 \documentclass[twocolumn,showpacs,preprintnumbers,amsmath,amssymb]{revtex4} 

 \usepackage{graphicx}
 \usepackage{dcolumn}
 \usepackage{bm}

\begin{document} 

 
 \title{Jamming at Zero Temperature and Zero Applied Stress:  the Epitome of 
 Disorder} 
 \author{Corey S. O'Hern} 
 \altaffiliation[Current address:]{ Department of Mechanical Engineering, 
 Yale University, New Haven, CT  06520-8284.} 
 \author{Leonardo E. Silbert} 
 \affiliation{Department of Chemistry and Biochemistry, UCLA, Los 
 Angeles, CA  90095-1569} 
 \affiliation{James Franck Institute, The University of Chicago, 
 Chicago, IL 60637} 
 \author{Andrea J. Liu} 
 \affiliation{Department of Chemistry and Biochemistry, UCLA, Los 
 Angeles, CA  90095-1569} 
 \author{Sidney R. Nagel} 
 \affiliation{James Franck Institute, The University of Chicago, 
 Chicago, IL 60637} 
 \date{\today} 

 \begin{abstract} 
 We have studied how 2- and 3- dimensional systems made up of particles 
 interacting with finite range, repulsive potentials jam ({\it i.e.,} 
 develop a yield stress in a disordered state) at zero temperature and 
 zero applied stress.  At low packing fractions $\phi$, the system is not 
 jammed and each particle can move without impediment from its 
 neighbors.  For each configuration, there is a unique jamming threshold, 
 $\phi_c$, at which particles can no longer avoid each other and the 
 bulk and shear moduli simultaneously become non-zero.  The 
 distribution of $\phi_c$ values becomes narrower as the system size 
 increases, so that essentially all configurations jam at the same 
 packing fraction in the thermodynamic limit.  This packing fraction 
 corresponds to the previously measured value for random close-packing.  
 In fact, our results provide a well-defined meaning for ``random 
 close-packing" in terms of the fraction of all phase space with inherent 
 structures that jam.  The jamming threshold, Point J, occurring at zero 
 temperature and applied stress and at the random close-packing density, 
 has properties reminiscent of an ordinary critical point.  As Point J is 
 approached from higher packing fractions, power-law scaling is found for 
 the divergence of the first peak in the pair correlation function and in 
 the vanishing of the pressure, shear modulus, and excess number of 
 overlapping neighbors.  Moreover, near Point J, certain quantities no 
 longer self-average, suggesting the existence of a length scale that 
 diverges at J.  However, Point J also differs from an ordinary critical 
 point: the scaling exponents do not depend on dimension but do depend on 
 the inter-particle potential.  Finally, as Point J is approached from 
 high packing fractions, the density of vibrational states develops a 
 large excess of low-frequency modes.  Indeed, at Point J, the density of 
 states is a constant all the way down to zero frequency.  All of these 
 results suggest that Point J may control behavior in its vicinity - 
 perhaps even at the glass transition.
 \end{abstract} 
 \pacs{81.05.Rm, 
 82.70.-y, 
 83.80.Fg 
 } 
 \maketitle

 \section{Introduction} 
 \label{intro} 
 The nature of the glass transition has been called probably ``the 
 deepest and most interesting unsolved problem in solid state 
 theory.''\cite{anderson} 
 The nature of granular materials has also 
   been said to lead to equally deep questions in statistical physics: 
 ``One might even say that the study of granular materials gives one a 
   chance to reinvent statistical mechanics in a new 
   context."\cite{kadanoff} Indeed, only a few years ago the state of 
   understanding of granular matter was compared to ``the level of 
   solid-state physics in 1930."\cite{degennes} There is no doubt that 
   there are hard and deep problems associated with both types of 
   systems and it may seem, at the outset, foolish to try to study both 
   problems simultaneously.  However, there have been significant 
   advances in both fields of study that indicate that these problems 
   are perhaps intimately related.  They both deal with amorphous 
   systems of particles in which the dynamics is perched precariously 
   near a transition between a flowing and a static state: that is, both 
   systems are close to a jamming threshold where all dynamics ceases. 
   One question that one can ask is whether there is something generic 
   about such transitions so that the freezing of a liquid into a glass 
   can profitably be compared to the arrest of a flowing granular 
   material, or a suspension, as external stresses 
   are reduced below the yield stress.  In other words, can one 
   study systems that can explore different 
   states either through thermal fluctuations or through externally 
   applied stresses, and search for unifying concepts that describe 
   their arrested dynamics as different aspects of a more general 
   ``jamming" behavior\cite{book}? 

   Our approach to this problem is to describe both glassy systems and 
   granular ones using the concept of a ``jamming phase diagram.''  In such 
   a diagram the ``phase boundary" marks the point where the response of 
   the system has become so sluggish as to make it appear solid on any 
   experimental time scale.  Using this framework, one can gain 
   insight into the relationship between athermal jamming and thermal 
   glass transitions and appreciate what are the control variables that 
   govern dynamical slowing down under many different conditions.  In 
   this paper, we describe simulations of a model liquid with 
   frictionless, 
   finite-range repulsive interactions. Because the potentials fall to 
   zero at some fixed finite radius, such a system may be a starting 
   point for understanding 
   macroscopic granular or colloidal systems as well as 
   liquids.  For such potentials, there is one special point, at zero 
   temperature and zero applied shear stress on the surface separating 
   the jammed and unjammed regions, which has exceptional and unique 
   properties.  The 
   goal of this paper is to elucidate some of the important properties 
   of this specific jamming transition in depth.  We have found that the 
   transition near this point has some aspects that resemble a critical 
   point and other properties that are not expected for a normal 
   second-order transition.  However, just as with a more conventional 
   critical point, there is the tantalizing possibility that it may 
   control the region around it and thereby govern the nature of the 
   entire jamming surface in the phase diagram. 

   We will first describe what is meant by jamming and what systems may 
   profitably be studied under this rubric.  We will then describe the 
   jamming phase diagram and show the important consequences that can be 
   drawn from it.  The nature of the transition at zero temperature and 
   zero applied shear stress will then be described to show why it is such an 
   important and unique transition. 

   \subsection{Systems that Jam} 
 \label{systems_that_jam} 
   Jamming occurs when a system develops a yield stress in a disordered 
   state \cite{book}.  In many cases, it is difficult to tell whether a 
   system has an infinite stress relaxation time (and hence a yield 
   stress), or whether it has a finite stress relaxation time that 
   exceeds the time scale of one's measurement.  An alternate definition 
   is therefore that jamming occurs when a system develops a stress 
   relaxation time that exceeds a reasonable experimental time scale in 
   a disordered state.  According to these definitions, many systems 
   jam.  Granular materials can flow when they are shaken or poured 
   through a hopper, but jam when the shaking intensity or pouring rate 
   is lowered \cite{granreview}.  Colloidal suspensions of particles are 
   fluid but jam when the pressure or packing fraction is raised 
   \cite{collreview}.  Foams and emulsions (concentrated suspensions of 
   deformable bubbles or droplets) flow when a large shear stress is 
   applied, but jam when the shear stress is lowered below the yield 
   stress \cite{foamreview}.  It should be emphasized here that granular 
   materials, foams and dense emulsions are athermal in the sense that 
   ordinary room-temperature thermal fluctuations are too insignificant 
   to allow the system to explore phase space.  However, for other 
   systems - typically those consisting of smaller particles, such as 
   molecular liquids - temperature plays an important if not dominant 
   role.  These liquids jam (if crystallization does not intervene 
   first) as temperature is lowered or density is increased--this is the 
   glass transition \cite{glassreview}.  There are a number of striking 
   similarities in the phenomenology of these different transitions. 
   Despite much effort, no significant static structural signature - as 
   opposed to a kinetic slowing down - of jamming has been observed 
   experimentally in any of these systems\cite{slowing_down}.  However, 
   we have proposed that such a signature can be observed in a quantity 
   initially measured for granular materials\cite{ohernpf}.  Another 
   similarity among the different systems is that the increase of the 
   stress relaxation time tends to be super-Arrhenius as a function of 
   the control parameter\cite{super}.  In addition, all systems show 
   kinetic heterogeneities near the onset of jamming, where particle 
   mobilities become heterogeneous in space and intermittent in time 
   \cite{mobilities}.  However, the parameters that control jamming 
   (temperature for the glass transition, applied shear stress for a 
   foam, packing fraction for a colloidal suspension) are so different that it 
   previously was difficult to see how to compare the jamming 
   transitions at a quantitative level. 

   \subsection{Jamming Phase Diagram} 
 \label{jamming_phase_diagram} 
\begin{figure} 
 \scalebox{0.5}{\includegraphics{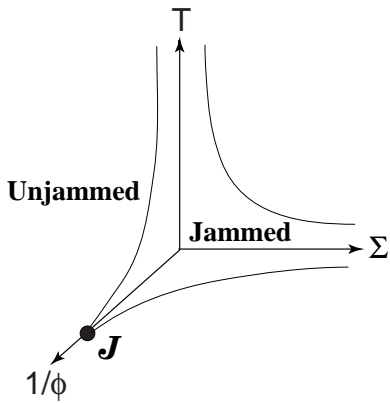}}%
   \caption{``Jamming phase diagram".  The jammed region, near the 
   origin, is enclosed by the depicted surface.  The point labeled ``J" is the 
   boundary of the jammed region at $T = 0$ and $\Sigma = 0$. Adapted from 
   Ref.~\cite{lucid}. 
   } 
   \label{jpd} 
   \end{figure}

   We proposed in Ref.~\cite{lucid} that different routes to kinetic 
   arrest 
   can be tied 
   together by a ``jamming phase diagram," shown schematically in 
   Fig.~\ref{jpd}.  The shape of the jamming surface may be different 
   for different systems.  The choice of axes is dictated by the parameters 
   that control the transition to jamming in the different systems, 
   namely temperature $T$, density or packing fraction $\phi$, and shear 
   stress $\Sigma$.  Note that $T$ and $\phi$ are traditional axes for 
   phase diagrams, but $\Sigma$ is not.  In the unjammed regime, the 
   system flows at nonzero $\Sigma$, so $\Sigma$ is a non-equilibrium 
   axis.  Why should there be such an axis in the jamming phase diagram? 
   One reason is that shear stress introduces fluctuations in the 
   unjammed regime by forcing the system to explore different packing 
   configurations.  Recent studies show that such fluctuations can be 
   described by an ``effective temperature" that has many of the 
   attributes of a true temperature 
   \cite{kurchan,berthier,makse,ono}.  Moreover, the dynamics of a 
   sheared system whose effective temperature is lowered toward jamming 
   are quantitatively similar to the dynamics of an equilibrium system 
   whose temperature is lowered toward the glass transition 
   \cite{langer,oherntbp}.  These results help to justify the existence 
   of shear stress as an axis on the phase diagram. 

   The ordinary phase diagram for the glass transition lies in the 
   vertical plane coming out of the page of Fig.~\ref{jpd}, namely the 
   $(1/\phi) - T$ plane.  At high packing fraction there is a transition 
   between a supercooled liquid and a glass that occurs at $T_{g}$. 
   (Although the relaxation times appear as if they will diverge close 
   to the transition line, it is impossible in practice to track their 
   increase past the times scales accessible to experiment.  Thus the 
   transition line, $T_{g}$, marks the position where the relaxation 
   time has reached some large threshold.  Its exact position may depend 
   to a small extent on the largest time that an experimentalist is 
   willing to run an experiment.  This definition corresponds to the 
   conventional one used for $T_{g}$ in glass-forming liquids.)  As the 
   packing fraction is lowered, $T_{g}$ normally decreases 
   \cite{kivelson}.  This glass-transition line is represented by the 
   curve separating the jammed (i.e. glass) and unjammed (i.e. liquid) 
   regions in the $(1/\phi) - T$ plane.  The ordinary phase diagram for 
   a foam or emulsion would be in the horizontal plane coming out of the 
   page, namely the $(1/\phi) - \Sigma$ plane of Fig.~\ref{jpd}.  At 
   fixed packing fraction, one must apply a shear stress higher than the 
   yield stress in order for the system to flow at an experimentally 
   measurable shear rate.  Thus, the yield stress as a function of 
   packing fraction is the curve that separates the jammed and unjammed 
   regions in this plane.  As the packing fraction decreases toward 
   close-packing, the yield stress typically decreases, as indicated in 
   Fig.~\ref{jpd}\cite{mason,stjalmes}. 

   Mode-coupling theorists suggested years ago that the colloidal glass 
   transition and molecular glass transition are the same despite the 
   fact that the control variables are different\cite{gotze}.  More 
   recently, mode-coupling theories have been extended to include shear 
   stress\cite{fuchs} or other control variables not derivable from 
   Hamiltonians \cite{berthier_theory}.  The jamming phase diagram suggests a 
   reason why different jamming transitions might be related, 
   independent of the validity of the mode-coupling approximation. 

   While it has long been recognized that temperature, packing fraction 
   and stress can all control the stress relaxation time, the concept of 
   the jamming phase diagram is a productive way to correlate jamming in 
   different amorphous systems.  The diagram implies that these three 
   control parameters are important to all systems, so that one can 
   study a single system as a function of all three variables.  The 
   diagram has proved to be a useful way to think about experiments, as 
   shown recently by Trappe, {\it et al.}\cite{trappe} on solidification 
   of attractive colloids.  It also explicitly suggests new experiments 
   to be done.  For example, it suggests that one should measure how the 
   relaxation time in a glass-forming liquid depends on applied stress. 
   It also suggests that the introduction of a temperature to an 
   otherwise jammed athermal system can help the system to flow.  That 
   is, temperature is a relevant variable for these transitions.  This 
   is, of course, in qualitative accord with the daily experience that 
   shaking an otherwise jammed material can reinitiate flow.  Perhaps 
   the most significant implication of the diagram is that the jammed 
   region might control the behavior nearby, and that this is why 
   different systems behave so similarly as they slow down on their 
   approach to the jammed state.

   \section{Jamming at Point J} 
 \label{jamming_at_pointj} 
   Perhaps the most daunting problem in studying any jamming transition 
   is that the jammed surface depicted in Fig.~\ref{jpd} is typically 
   not sharp, and is defined by the system's relaxation time exceeding 
   experimental time scales.  
 However, there is one point on the jamming phase diagram that is 
   well-defined\cite{ohernJ}, namely, the point labeled ``J" in 
   Fig.~\ref{jpd}. 
   This point exists at zero 
   temperature and zero applied shear stress for systems with repulsive, 
   frictionless, finite-range potentials.  This section is devoted to 
   the special properties of 
   Point J. 

   \subsection{Method} 
 \label{method} 
   To explore Point J, we have studied potentials of the following form: 
 \begin{equation} 
 \label{potentialdef} 
 V(r_{ij}) = \left\{ \begin{array}{cc} 
 \epsilon (1-r_{ij}/\sigma_{ij})^\alpha/\alpha & {\rm for}~r_{ij} 
 < \sigma_{ij} \\ 
 0 & {\rm for}~r_{ij} \ge \sigma_{ij} 
 \end{array} 
 \right. 
 \end{equation} 
   where $\epsilon$ is the characteristic energy scale of the 
   interaction, $r_{ij}$ is the separation between the centers of 
   particles $i$ and $j$ and $\sigma_{ij}$ is the sum of the radii of 
   particles $i$ and $j$.  We study three different potentials, namely, 
   $\alpha = 2$ for repulsive harmonic springs, $\alpha = 3/2$ for 
   repulsive nonlinear springs that are harder than harmonic 
   springs, and $\alpha = 5/2$ for 
   repulsive Hertzian interactions that are softer than harmonic 
   springs\cite{hertzian}.  It is important to 
   note that the interactions are finite in range--particles do not 
   interact unless they overlap. 
   Potentials of this form were motivated by granular materials 
   where particles have a well-defined diameter 
   and do not interact except for a strong repulsive force that keeps 
   the particles from deforming too much.  In our two-dimensional ($2d$) 
   simulations we have used $50$-$50$ mixtures of particles with a size 
   ratio of $1.4$ in order to prevent crystallization\cite{perera,speedy1}.  
   The 
   diameter of the smaller particle is denoted by $\sigma$.  In three 
   dimensions ($3d$) we have studied the same bidisperse mixture as well as 
   monodisperse systems with particle 
   diameter $\sigma$.  We have studied the finite-size effects by 
   varying the number of particles in the sample between $4 < N < 4096$ 
   in $2d$ and $3d$. 

   Of crucial importance is the protocol for the creation of 
   configurations at $T = 0$ and a given packing fraction, $\phi$.  To 
   obtain such states, we start each simulation with a fixed number of 
   particles, $N$, with the particle positions chosen completely at 
   random (this corresponds to $T = \infty$) within a square or cubic box with 
   side length $L$ and periodic boundary conditions.  Starting with 
   randomly-generated $T=\infty$ states 
   guarantees that we sample all phase space equally.  We then bring the 
   system to the nearest potential-energy minimum by constantly moving 
   downward on the potential energy surface.  We do this using 
 conjugate-gradient 
   techniques \cite{numrec}.  Each conjugate gradient energy 
   minimization is terminated when one of the following two stopping 
   criteria is satisfied: 1) the total potential energy per particle 
   satisfies $V/N < 10^{-16}$ (this corresponds to a very small pressure, 
   $p < 10^{-10})$ or  2) $V/N$ for 
   successive iterations deviates by less than $10^{-15}$.  This 
   procedure brings the system extremely close to $T=0$.  Note 
   that this procedure is identical to that for finding the ``inherent 
   structures'' of the $T=\infty$ states \cite{stillinger}. 

    In addition to studying the $T=0$ states generated by the protocol 
   described above, we explore their properties by perturbing them 
   slightly.  We compress them, decompress them, or apply shear strains. 
   After each infinitesimal perturbation, we can again employ the 
   conjugate-gradient technique.  Since this technique takes the system 
   to the bottom of its local potential well, the quantities we measure 
   in this way are related to the static, or infinite-time ($t = 
   \infty$), response (the static bulk or shear moduli, $B_{\infty}$ or 
   $G_{\infty}$) of the configurations.  We have also measured the $t = 
   0$ moduli, $B_{0}$ and $G_{0}$, by measuring the response to a 
   perturbation immediately after it has been applied (before minimizing 
   the energy by the conjugate gradient technique). 
   The shear and bulk moduli are obtained 
 by measuring the response of the pressure tensor\cite{allen} 
 \begin{equation} 
 \label{pressure_tensor} 
 p_{\alpha \beta} =   -L^{-d} \sum_{i>j} r_{ij\alpha} \frac{r_{ij\beta}} 
 {r_{ij}} \frac{dV}{dr_{ij}} 
 \end{equation} 
 to shear and compression perturbations, where $r_{ij\alpha}$ is the 
 $\alpha$-component of ${\vec r}_{ij}$ and $d$ is the dimensionality of 
 the system.  To measure the bulk modulus we 
 calculate $B=\phi dp/d\phi$, where the pressure is $p = \sum_{\alpha} 
 p_{\alpha \alpha}/d$.  To measure the shear modulus, we calculate 
 $G= d \Sigma/d\gamma$, where $\Sigma = -p_{xy}$, after applying a shear 
 shear strain in the $x$-direction 
 with a strain gradient in the $y$-direction.  The pressure $p$, stress 
 $\Sigma$, bulk modulus $B$, and shear modulus $G$ are measured in 
 units of $\epsilon/\sigma^d$, lengths are measured in units of 
 $\sigma$, and timescales or inverse frequencies are measured in units
 of $\sigma \sqrt{m/\epsilon}$ where all particles have equal mass $m$.

   \subsection {J represents the onset of jamming for a single configuration} 
 \label{onset} 

   It is important to note that each initial $T = \infty$ state can
   yield a different value of the packing fraction, $\phi_{c}$, 
   where the pressure
   and potential energy first becomes nonzero.  Despite
   this ambiguity about the value of the threshold $\phi_{c}$, 
   we find that there are robust results when we
   measure properties as a function of $\phi-\phi_{c}$, including
   scaling laws, that appear to be the same for all initial
   configurations.  In Sec.~\ref{infinite_limit}, we will examine the
   nature of the distribution of these values of $\phi_c$.  In this
   subsection we will show that it is possible to locate a
   well-defined onset of jamming, $\phi_{c}$, for each initial state.

   To test whether a given $T = 0$ state is jammed or not, two separate 
   criteria must be met: a jammed state must have a non-zero static 
   (i.e., infinite-time) value of both the bulk 
   modulus and the shear modulus.  As we show below, for each state that 
   we have studied, the static bulk and static shear moduli approach 
   zero at the same density, $\phi_{c}$.  Thus, $\phi_{c}$ specifies 
   the onset of jamming for each state. 

   At $T=0$ and $\Sigma=0$, no two particles can interact if the density 
   is low enough.  If two particles were to overlap, their repulsive 
   potentials would simply push them apart during the conjugate gradient 
   energy minimization process until they no longer touched.   Since there 
   is neither thermal energy nor shear stress to compete with the 
   particles' potential energy, they will never be forced back into 
   contact.  Thus, at sufficiently low densities there are no particle 
   overlaps and the final potential energy, $V$, and the pressure, $p$, 
   are both zero so that the system has a zero static bulk modulus.  At 
   the threshold packing fraction, $\phi_c$, particles just come into 
   unavoidable contact since there is no longer enough free space to 
   allow them to move apart.  As the system is compressed further, the 
   particles overlap, the energy and pressure are nonzero and the bulk 
   modulus is nonzero because the pressure increases upon compression. 

   \begin{figure} 
 \scalebox{0.5}{\includegraphics{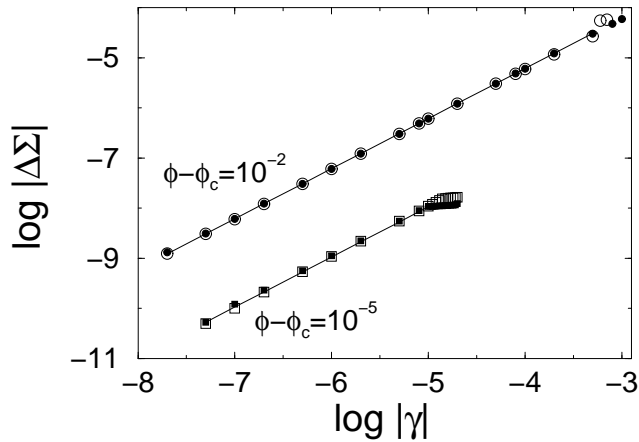}}%
   \caption{The infinite-time stress $\Delta \Sigma =
   \Sigma(\gamma) - \Sigma(0)$ following an applied
   shear strain $\gamma$.  The resulting stress-strain curve is linear
   for sufficiently small strains and independent of the sign of the
   strain.  Open (filled) symbols indicate negative (positive)
   strains.  These curves were generated using $3d$ monodisperse
   systems ($N=512$) with harmonic repulsions.  Circles and squares
   represent systems with packing fractions $\phi-\phi_c=10^{-2}$ and
   $10^{-4}$, respectively.  The solid lines have slopes equal to $1$.
   The shear modulus, yield stress, and yield strain (where stress
   versus strain becomes nonlinear) tend to zero as $\phi$ approaches
   $\phi_c$, where $\phi_c$ is the onset of jamming for a given
   configuration.}
   \label{shearmod} 
   \end{figure} 

   For each initial $T=\infty$ state, we first obtain a $T = 0$ state 
   using conjugate gradient minimization.  For that $T=0$ state, we 
   measure a precise value of $\phi_c$, as follows: If the configuration 
   has zero pressure, we compress the system (by increasing the size 
   of each particle by the same fixed fraction) in very small steps, 
   applying conjugate-gradient energy minimization after each step, 
   until the pressure becomes nonzero at $\phi_{c}$. Conversely, if the 
   configuration has a non-zero pressure, we decompress the system 
   in small steps, applying conjugate-gradient energy minimization after 
   each step, until the pressure reaches zero at $\phi_{c}$. 
   We insure that 
   the system does not cross over any energy barriers during these 
   procedures by compressing (or decompressing) in successively smaller 
   increments.  As the 
   density variation is made finer and finer, we thus make sure we end 
   up in precisely the same configuration for all the particles 
   independent of the size of the increment.  Increments 
   were in the range $\Delta \phi=[10^{-6},10^{-4}]$, with 
   smaller increments used for smaller systems and systems closer 
   to $\phi_c$.

   At each packing fraction, we measure the static shear modulus, 
   $G_\infty$, by applying a very small shear strain, minimizing the 
   energy with the conjugate gradient technique, and measuring the final 
   induced stress. (Again, we insure that no energy barriers are crossed 
   by applying successively smaller increments of shear strain.  The 
   strain increments were in the range $[5\times 10^{-8},10^{-5}]$ with 
   smaller increments used for smaller systems and systems closer to 
   $\phi_c$.)  The shear modulus is calculated by measuring the linear 
   relation between stress and strain, as shown in 
   Fig.~\ref{shearmod}.  

\begin{figure} 
   \scalebox{0.5}{\includegraphics{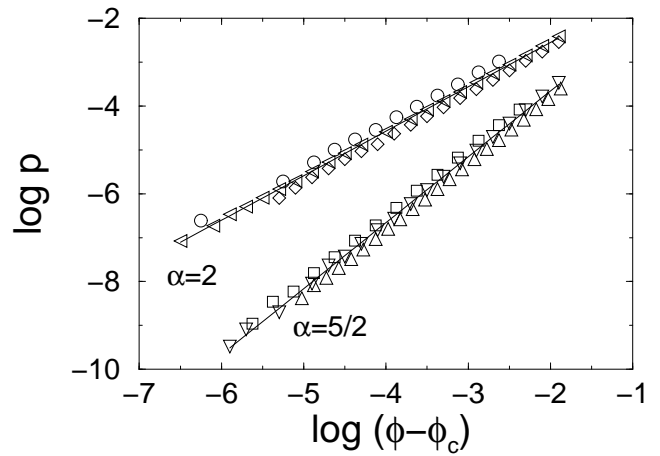}}%
   \caption{Upper curves: Pressure $p$ vs. 
   $\phi-\phi_{c}$ for $3d$ 
   monodisperse (circles), $3d$ bidisperse (diamonds), and 2d bidisperse 
 (leftward triangles) systems with harmonic repulsions ($\alpha=2$).  
 The solid 
 line has slope of $2.0$.  Lower curves: 
   $p$ vs. $\phi-\phi_{c}$ for 3d 
   monodisperse (squares), 3d bidisperse (upward triangles), and 2d bidisperse 
 (downward triangles) systems with Hertzian repulsions ($\alpha=5/2$). 
 The solid 
 line has a slope of $2.5$.  These symbols for the different systems 
 are used throughout the text.  $N=1024$ ($N=512$) particles were used 
 for the $2d$ ($3d$) systems.} 
   \label{pressure} 
   \end{figure}

 Fig.~\ref{pressure} shows the results for the pressure $p$
 as a function of $\phi-\phi_c$ for monodisperse systems in
 3-dimensions using both harmonic ($\alpha = 2$) and Hertzian
 ($\alpha= 5/2$) potentials.  We also include our earlier results for
 bidisperse systems in $2$ and $3$ dimensions using those same two
 potentials\cite{ohernJ}.  We find that the data for $p$ as a
 function of $\phi-\phi_{c}$ collapse onto a single curve for
 different initial states (each set of points corresponds to data from
 5 different states).  Thus, although each initial state has a
 different value of $\phi_{c}$, all states behave the same way as a
 function of $\phi-\phi_{c}$ when compressed above $\phi_{c}$.

\begin{figure} 
   \scalebox{0.5}{\includegraphics{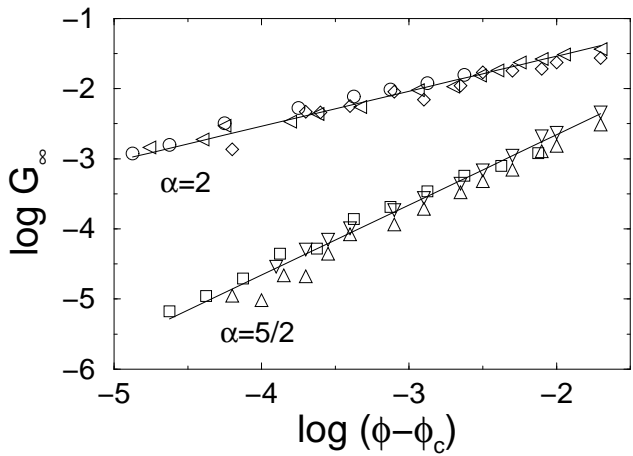}}%
   \caption{Upper curves: Static shear modulus $G_{\infty}$ vs. 
   $\phi-\phi_{c}$ for 
   3d monodisperse (circles), 3d bidisperse (diamonds) and 
   2d bidisperse (leftward triangles) systems with 
   harmonic repulsions ($\alpha=2$).  The solid 
 line has a slope of $0.5$.  Lower curves: $G_{\infty}$ vs. $\phi-\phi_{c}$ 
   for 3d monodisperse (squares), 3d bidisperse (upward triangles), and 
   2d bidisperse 
   (downward triangles) systems with Hertzian potentials ($\alpha=5/2$). 
 The solid line has a slope of $1.0$.  $N=1024$ ($N=512$) particles were 
 used for the $2d$ ($3d$) systems.} 
   \label{shearmodinfty} 
   \end{figure}

   In Fig.~\ref{shearmodinfty}, we show the static shear modulus,
   $G_{\infty}$, for the same initial states as shown for the
   pressure.  Again, we find that data for different initial states
   collapse on a single curve when $G_{\infty}$ is plotted against
   $\phi-\phi_{c}$.  Note that $\phi_{c}$ was determined by where the
   {\it pressure} approaches zero, not by where the static shear
   modulus first approaches zero.  Thus, Figs.~\ref{pressure} and
   ~\ref{shearmodinfty} show that the static shear modulus,
   $G_{\infty}$, and the pressure, $p$ (and therefore, the
   static bulk modulus, $B_{\infty}$, as well), approach zero at the
   same packing fraction, $\phi_{c}$, to a precision of better than
   $2$ parts in $10^{5}$ for the monodisperse systems.  Each state
   develops a bulk modulus and shear modulus at the same packing
   fraction.  This is true for all polydispersities, dimensionalities
   and potentials studied.  Thus, {\it $\phi_{c}$ truly marks the
   onset of jamming for a given initial state.}

   Note that in measuring the static shear modulus, we apply a shear 
   stress in a given direction.  Although we have shown that every state 
   studied can withstand a shear stress in that direction for 
   $\phi>\phi_{c}$, it is not obvious from these measurements that every state 
   can withstand a shear stress in {\it any} arbitrary direction.  To 
   address this, we have studied the eigenvalues of the dynamical 
   matrix\cite{dynamical} for our $T=0$ configurations with harmonic 
   repulsions. 
   We find that at least for 
   $\phi-\phi_{c} \ge 10^{-6}$, the only zero-frequency modes correspond to 
   isolated clusters of
   ``rattlers,'' i.e. particles that do not overlap with any other 
   particles and to uniform translations of the entire system.  The lack 
   of any nontrivial zero-frequency modes shows 
   unambiguously that the system can withstand a shear stress in all 
   directions.  We discuss the statistics of rattlers in greater detail 
   in Section ~\ref{isostatic_section} and the properties of the 
   dynamical matrix in more detail in Section 
   ~\ref{density_of_states_section}. 

   \subsection{Onset of jamming is sharp in the limit of infinite 
   system size} 
 \label{infinite_limit} 

\begin{figure} 
   \scalebox{0.5}{\includegraphics{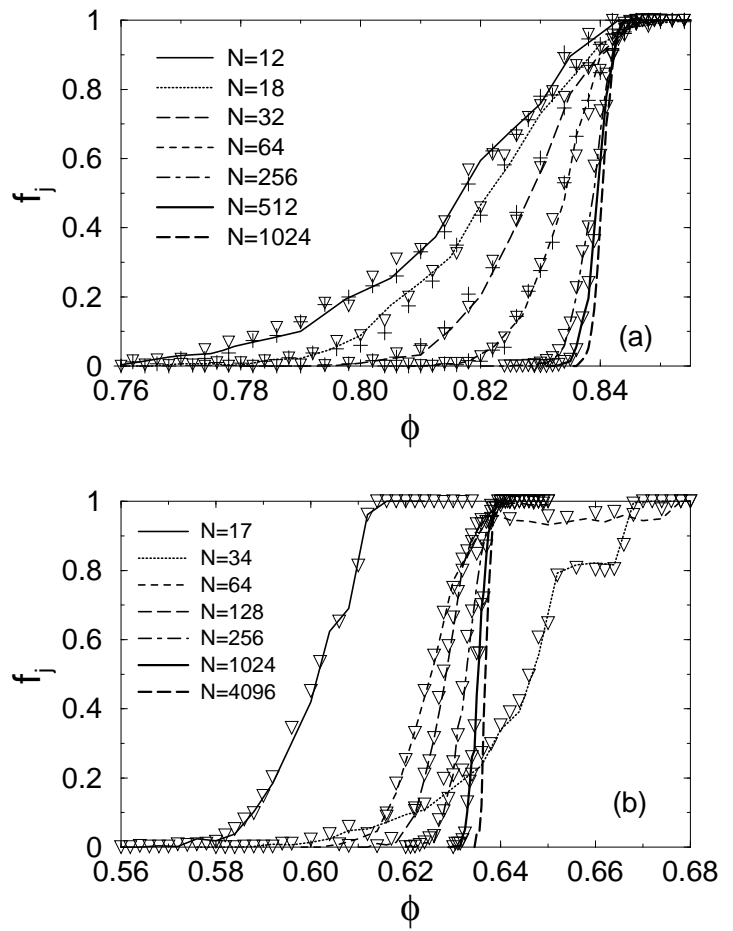}}%
   \caption{Fraction $f_j$ of jammed 
   states as a function of $\phi$ for (a) $2d$ bidisperse systems 
   and for (b) $3d$ monodisperse systems with harmonic and Hertzian 
   repulsions.  In (a) and (b), 
   the lines (downward triangles) represent potentials with $\alpha=2$ 
   ($\alpha=5/2$).  $f_j$ for 2d bidisperse 
   systems with 
   $\alpha=3/2$ are also shown in (a) using plus symbols.  Each curve 
   represents a different system size $N$.} 
   \label{Pjam} 
   \end{figure}

   In the last subsection, we showed that different initial random ($T = 
   \infty$) states have inherent structures ($T=0$ states) that jam at 
   different threshold values, $\phi_c$.  Here we measure 
   the distribution of jamming thresholds.  For each system size $N$ and 
   packing fraction $\phi$, we start with at least $500$ ($100$ for the 
   largest system sizes) random ($T = \infty$) configurations and use 
   the conjugate gradient method to quench each configuration infinitely 
   rapidly to $T = 0$.  We then find the fraction of these final states 
   that are ``jammed", {\it i.e.}, that have a finite pressure and 
   static shear 
   modulus.  The resulting 
   fraction $f_j$ of jammed states is shown as a function of $\phi$ in 
   Fig.~\ref{Pjam}(a) for a two-dimensional bidisperse system and in 
   Fig.~\ref{Pjam}(b) 
   for a three-dimensional monodisperse system with harmonic 
   repulsions.  Similar graphs were shown for three-dimensional 
   bidisperse systems with harmonic repulsions in Ref.~\cite{ohernJ}. 

   In measuring these distributions, the system remains at one fixed, 
   well-defined density since we do {\it not} dilate 
   or shrink the particles.  Also, during the quench itself, 
   there are no dynamics.  The system only travels on the potential 
   energy surface and descends via the most rapid route to the nearest 
   local potential-energy minimum.  This distribution is therefore {\it not} 
   a function of the dynamics used in obtaining the final configurations 
   but depends only on the fixed potential energy landscape.  By starting 
   with $T=\infty$ states we are sampling configuration 
   space uniformly.  Thus, the result shown in Fig.~\ref{Pjam}(a) and (b) is a 
   measure of the total fraction of configuration space (i.e., the 
   probability) that belongs in the basins of attraction of final 
   configurations that are jammed. 

     \begin{figure} 
     \scalebox{0.47}{\includegraphics{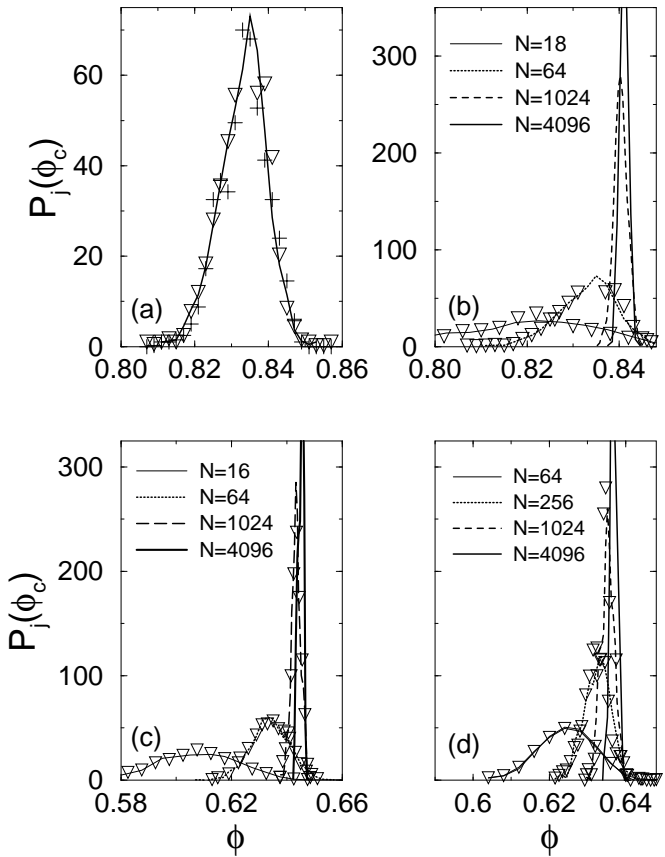}}%
     \caption{(a) Distribution of jamming thresholds $P_j(\phi_c)$ for
     a 2d bidisperse system with $N=64$ for the three different
     potentials studied ($\alpha=3/2$, $2$, and $5/2$).  $P_j(\phi_c)$
     for (b) $2d$ bidisperse systems, (c) $3d$ bidisperse systems, and
     (d) $3d$ monodisperse systems with harmonic and Hertzian
     potentials for various system sizes.  In (a)-(d), the pluses,
     lines, and downward triangles represent potentials with
     $\alpha=3/2$, $2$, and $\alpha=5/2$, respectively.  The
     distributions for small 3d monodisperse systems ($N<64$) were not
     shown in (d) because we wanted to emphasize the monotonic
     behavior of the peak in $P_j(\phi_c)$ at large $N$.}
     \label{distribution} 
     \end{figure} 
     
   Fig.~\ref{Pjam}(a) and (b) show that the fraction of jammed states depends 
   sensitively on system size.  For the 2d bidisperse system 
   (Fig.~\ref{Pjam}(a)), the 
   curves progressively sharpen with increasing $N$, eventually 
   approaching a vertical jump.  The 3d 
   monodisperse system (Fig.~\ref{Pjam}(b)) shows similar 
   behavior for $N>64$.  For smaller values of $N$, there is enough 
   partial crystallization to produce additional structure in the 
   curves. 

   We calculate the distribution of jamming thresholds
   $P_{j}(\phi_{c})$ by differentiating the data in Fig.~\ref{Pjam}
   with respect to $\phi$.  We find that the distributions are
   insensitive to the inter-particle potential used.  This is
   illustrated in Fig.~\ref{distribution}(a) for 2d bidisperse systems
   at fixed system size $N=64$.  In this figure, we overlay the
   distributions for $\alpha=5/2$ (Hertzian repulsions; downward
   triangles) and $\alpha=3/2$ (plus symbols) on top of the $\alpha=2$
   (harmonic; solid lines) distributions.  In
   Figs.~\ref{distribution}(b)-(d), we overlay the distributions for
   $\alpha=5/2$ on top of those for $\alpha=2$ for all systems studied
   ($2d$ bidisperse, $3d$ bidisperse, and $3d$ monodisperse) at
   several system sizes $N$.  Within numerical error, the different
   potentials yield identical distributions at each $N$.

\begin{figure} 
   \scalebox{0.5}{\includegraphics{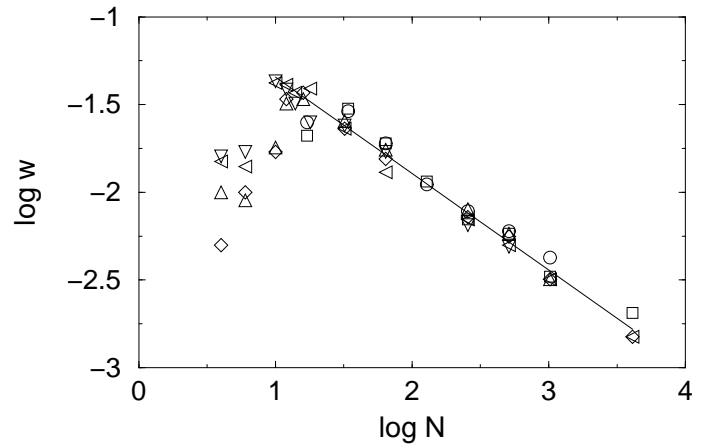}}%
     \caption{Width of the distribution of jamming thresholds $w$ vs. the number 
     of particles $N$ for $2d$ 
     bidisperse, $3d$ bidisperse, and $3d$ monodisperse systems with 
     harmonic and Hertzian potentials.  The solid curve has a slope of 
     $-0.55$.  The symbols have the same meaning as in 
     Fig.~\ref{pressure}.   } 
     \label{width} 
     \end{figure}

   Fig.~\ref{distribution} also shows that it is unlikely that a
   jamming threshold $\phi_{c}$ will be found at very low packing
   fraction, where almost all states are unjammed, or at very high
   packing fraction, where almost all states are already jammed.  For
   small systems, the distributions are broad; as $N$ increases, they
   become sharper and taller.  To quantify the change of the
   distributions with system size, we extract the full width at half
   maximum of the distribution, $w$, for each $N$.  The results are
   plotted in Fig.~\ref{width} and are not monotonic in $N$.  At very
   small $N$, there are only a few distinct configurations available
   to a static packing, so the distribution of jamming thresholds is
   narrow.  The width grows with increasing $N$ to a maximum (near $N
   = 10$ for bidisperse systems and near $N=30$ for $3d$ monodisperse
   systems).  Above this value, the width decreases with increasing
   $N$.  At the system size where the distributions are widest, 
   there is a reasonable probability of systems jamming at
   packing fractions as low as roughly $\phi=0.80$ in $2d$ bidisperse
   systems and $\phi=0.58$ in $3d$ bidisperse and monodisperse
   systems.  Perhaps this is a coincidence, but it is
   interesting that the value in 3d corresponds to previous
   estimates of ``random loose-packing'' from experiments\cite{onoda}.
   It has been reported that hard particle methods (methods that
   strictly prohibit particle overlap) can produce jammed states with
   packing fractions that are much lower than the peak in the distribution of
   jamming onsets\cite{speedy2}.  However, we have carried out similar
   hard particle simulations and find that these low-$\phi$ states are
   not jammed according to our definition given above.  Instead, these
   states are nearly unjammed and fall apart when they are slightly
   compressed or sheared.
   
   In the large $N$ regime, 
   Fig.~\ref{width} shows that the full width at half maximum of the 
   distribution scales as 
   \begin{equation} 
        w = w_0 N^{-\Omega} \label{widthpower} 
   \end{equation} 
   with $\Omega = 0.55 \pm 0.03$ and $w_0=0.16 \pm 0.04$ for all of the 
   systems studied.  This implies that as $N$ diverges, the width 
   approaches zero and the distribution of jamming thresholds approaches 
   a $\delta$-function.  In other words, {\it in the thermodynamic limit, 
   essentially all of phase space jams at the same packing fraction, 
   $\phi^{*}$}.  This means that Point J in the jamming phase diagram is 
   well-defined as the onset of jamming.

 \subsection{Point J is random close-packing in an infinite-size system} 
 \label{rcp_section} 

 Our results are relevant to hard-sphere systems because the $T=0$ 
   configurations obtained by this protocol are allowed hard-sphere 
   configurations if none of the particles overlap.  Thus, at 
   sufficiently low $\phi$, the conjugate gradient minimization 
   technique will invariably yield allowed hard-sphere states. 
   Our protocol yields 
   special insight into the nature of random close-packing, a highly 
   reproducible but heretofore somewhat vaguely-defined state. 

\begin{figure} 
 \scalebox{0.5}{\includegraphics{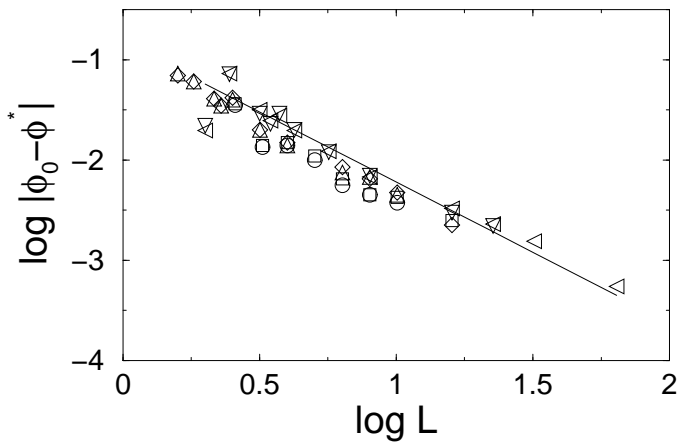}}%
   \caption{Deviation of the peak in the distribution of jamming thresholds 
   from its asymptotic value $|\phi_0 -\phi^*|$ vs. $L$ for $2d$ bidisperse, 
   $3d$ bidisperse, and $3d$ monodisperse systems with harmonic and 
   Hertzian potentials.  The solid curve has a slope of $1/\nu=1.40$. 
   The symbols have the same meaning as in 
     Fig.~\ref{pressure}. } 
   \label{peak} 
   \end{figure}

   We make the connection to random close-packing by asking what is the 
   limiting $N \rightarrow \infty$ value of the jamming 
   threshold, $\phi^{*}$.  We calculate it by extrapolating the {\it 
   peak positions}, $\phi_{0}$, of the distributions shown in 
   Fig.~\ref{distribution} with respect to the system size.  In 
   Fig.~\ref{peak}, we plot the deviation of $\phi_0$ from $\phi^*$ as a 
   function of $L \equiv N^{1/d}$, where $d$ is the dimensionality. 
   The peak position approaches its limiting asymptotic value as a power 
   law in $L$: 
   \begin{equation} 
        \phi_{0}-\phi^{*} = \delta_0 L^{-1/\nu} \label{peakpower} 
   \end{equation} 
   By fitting to this form, we obtain $\nu = 0.71 \pm 0.08$ and 
   $\delta_0= 0.12 \pm 0.03 $ for all systems studied. 
   Previously\cite{ohernJ}, we obtained $\phi^{*}$ for bidisperse 
   systems in 2 and 3 dimensions.  For monodisperse 3-dimensional 
   systems, we now find 
 \begin{equation} 
 \label{rcp} 
 \phi^* =  0.639 \pm 0.001 
 \end{equation} 
   We find that $\phi^*$ does not vary with potential; this 
   follows from our result that the distributions of jamming thresholds 
   are independent of potential ($\alpha=3/2$, $\alpha=2$, $\alpha=5/2$) 
   within the uncertainty of the measurement. 
   Note that the value of $\phi^*$ in Eq.~\ref{rcp} for monodisperse
   3-dimensional systems is very close to what has been reported for
   random-close packing.  Our value should be compared to other recent
   estimates of random close packing, $\phi_{rcp} \approx
   0.64$\cite{berryman,torquatoprl}.  This similarity is not a
   coincidence.  

   Random close-packing cannot be defined in a
   mathematically precise way because the terms ``random'' and
   ``close-packed'' are at odds with one another\cite{torquatoprl}.
   Because the close-packing density of an fcc packing is
   $\pi/\sqrt{18} \approx 0.74 > 0.64$, one can always make the system
   more highly close-packed (but less random) by introducing some
   degree of crystalline order.  How ``random'' the system should be
   versus how ``close-packed'' it should be is arbitrary.  Torquato,
   {\it et al.} therefore propose another term, ``maximally random
   jammed state.''  By ``jammed,'' they mean that any particle or set
   of particles cannot be translated relative to any of the rest of
   the particles in the system, and by ``maximally random,'' they
   suggest a definition based on minimization of order parameters
   characterizing the extent of crystalline order, bond orientational
   order, etc.\cite{torquatopre}.

   Here we suggest an alternate interpretation for random close-packing 
   using the language of a ``maximally random 
   jammed state'' but with different meanings attached to ``maximally 
   random'' and ``jammed.''  In our case, the value $\phi^{*}$ is 
   obtained by extrapolating the peak of the distribution of jamming 
   thresholds to infinite system size.  The peak of the distribution 
   corresponds to the packing fraction with the maximum fraction of 
   phase space (i.e., the {\it maximum entropy}) that belongs to the 
   basin of attraction of jamming thresholds in the 
   thermodynamic limit.  We therefore propose 
   that another way to 
   define ``maximally random'' is by where the entropy of initial states 
   is a maximum, and that another way to define ``jammed'' is by the 
   disappearance of zero-frequency modes of the dynamical matrix (with 
   the exception of isolated clusters of rattlers).  This definition has 
   the advantage of avoiding the order parameter description, which will 
   always be subject to uncertainty since one never knows if one has 
   calculated the proper order parameter.  It also provides a cleaner 
   definition of the word ``jammed,'' since it depends on nature of 
   zero-frequency modes of the dynamical matrix.  If one is testing 
   whether a system is jammed by shifting particles, it is unlikely that 
   one will hit on the exact combination of particle shifts that is 
   characterized by the eigenvector of a zero-frequency mode.  Finally, we note 
   that our finding that virtually all initial states jam at the same value, 
   $\phi^{*}$, in the thermodynamic limit may explain why the value of 
   random close-packing is so robust despite the fact that it has not 
   been well-defined in the past.  Although regions of 
   the system can crystallize, such states are extremely rare and 
   therefore unlikely to be observed for sufficiently large systems. 

   The above definition of random close-packing, or the ``maximally random 
   jammed state,'' is completely well-defined for soft, finite-ranged repulsive 
   potentials.  What can be said about hard spheres? We can 
   approach the hard-sphere limit by making the potential harder and 
   harder--that is, by making the exponent in the potential, $\alpha$, (see 
   Eq.~\ref{potentialdef}) 
   approach 0.  Measuring  $\phi^*$ as a function of $\alpha$ will 
   then produce a 
   limiting hard-sphere, value for random close packing.  Note that our 
   results for $\phi^{*}$ are the same, within measurement error, for 
   $\alpha=3/2$, $\alpha=2$ (harmonic) and $\alpha=5/2$ (Hertzian).  Thus, the 
   value of $\phi^{*}$ is insensitive to $\alpha$, 
   suggesting that the hard-sphere limit of $\phi^{*}$ is the 
   same as the value we have given in Eq.~\ref{rcp}.  Of course, it is 
   not clear that the hard-sphere limit is well-defined; different ways 
   of taking the hard-sphere limit may lead to different results. 
   If that is indeed the case, we would argue that hard spheres are 
   a singular limit and thus unphysical.  One should therefore concentrate 
   on softer potentials for which unambiguous definitions can be constructed. 

   Another way that has often been employed to study hard-sphere 
   configurations near random close-packing is to conduct density ramps. 
   For example, in the Lubachevsky-Stillinger 
   algorithm\cite{lubachevsky}, a hard-sphere system at low packing 
   fraction is suddenly compressed (by increasing the radii of all the 
   particles at some fixed rate) to a higher packing fraction.  In the 
   limit of infinite quench rate, one finds that the system jams at a 
   random close-packing density.  One advantage to our protocol for 
   systems with softer but still finite-ranged repulsive potentials is 
   that, since the density is always held constant, we can quench the 
   system to the final state within a fixed energy landscape.  In the 
   Lubachevsky-Stillinger algorithm, the energy landscape changes 
   throughout the density ramp because the density necessarily varies 
   throughout the procedure. 

   One of the strengths of our procedure is that dynamics has no role. 
   If we introduce dynamics by quenching the temperature of the system 
   at some finite rate, we bias the distributions of jamming thresholds 
   toward higher values of $\phi$.  These distributions no longer 
   represent features only of the potential energy surface but now 
   also depend on dynamics through the quench rate.  By contrast, our 
   distributions are solely a geometric property of the potential energy surface. 

   \subsection{Point J is an isostatic point} 
 \label{isostatic_section} 

\begin{figure} 
   \scalebox{0.5}{\includegraphics{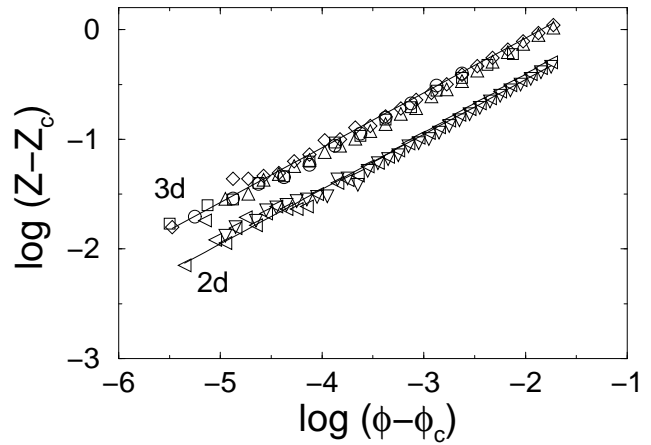}}%
   \caption{Upper curves: Excess number of contacts per particle $Z-Z_c$ vs. 
   $\phi-\phi_c$ 
   for $3d$ systems: monodisperse, harmonic (circles); monodisperse, 
   Hertzian (squares); bidisperse, harmonic (diamonds); bidisperse, 
   Hertzian (upward triangles).  Lower curves:  $Z-Z_c$ versus $\phi-\phi_c$ for 
   $2d$ systems: bidisperse, harmonic (leftward triangles) and bidisperse, 
   Hertzian (downward triangles).  $N=1024$ ($N=512$) particles were 
   used for the $2d$ ($3d$) systems. The symbols have the same meaning as in 
   Fig.~\ref{pressure}.} 
   \label{Z} 
   \end{figure}

   An isostatic configuration is defined by having the number of 
   contacts in the system, $NZ/2$, equal to the number of force balance 
   equations \cite{shlomo}, where $Z$ is the average number of contacts 
   per particle.  When this occurs, there is a unique 
   solution for the forces between particles in a static packing, 
   because the number of equations equals the number of unknowns.  For 
   purely repulsive, frictionless systems of spherical particles, the 
   number of force balance equations is $N d$ so the 
   isostatic condition is $Z=2d$, where $d$ is the 
   dimensionality of the system.  We find\cite{ohernJ} that 
   there is a discontinuous jump in $Z$ at the jamming threshold, 
   $\phi_{c}$, of a given state.  For $\phi =\phi_{c}^{-}$, there are no 
   overlapping neighbors, $Z = 0$, while for $\phi=\phi_{c}^{+}$ there 
   are $Z_{c}$ overlapping neighbors.  The value of $Z_{c}$ can be obtained 
   by measuring $Z$ at 
   values just above $\phi_{c}$, as shown in Fig.~\ref{Z}.  The straight 
   lines in the plots are fits to the data of the form 
   \begin{equation} 
        Z-Z_{c} = Z_{0} (\phi-\phi_{c})^{\zeta}, \label{Zpower} 
   \end{equation} 
   where $\zeta=0.50 \pm 0.03$ for all potentials, dimensions and 
 polydispersities studied.

   As mentioned in Sec.~\ref{onset}, approximately $5\%$ of the particles 
   are ``rattlers'' with no contacts at all, which do not contribute to 
   the connected network. 
   If we exclude the rattlers 
   (so that we are 
   only studying properties of the connected network) and assume 
   $\zeta=0.5$, then we obtain precise values for $Z_c$, listed in 
   Table ~\ref{numbers}. 
   These results are consistent with $Z_{c}=2d$ in all cases, implying that 
   the jamming threshold is an isostatic point.  In the thermodynamic 
   limit, $\phi_{c} \rightarrow \phi^{*}$, so Point J is 
   an isostatic point.  Note that our results for $Z$ show that Point J 
   is the {\it only} point at which the packing is isostatic; above 
   $\phi^{*}$, we find $Z>2d$ so additional equations (the constitutive 
   relations for the particles, which depend on the potential used) are 
   needed to solve for the forces between particles. 

   A more stringent condition for isostaticity is that the connected 
   network ({\it i.e.} all particles in the system excluding rattlers) has no 
   zero-frequency modes.  As discussed in Sec.~\ref{onset}, we have looked 
   for zero-frequency modes in packings above $\phi_{c}$, and have tested 
   configurations with packing fractions as little as $10^{-6}$ above 
   $\phi_{c}$.  For all configurations tested, we have seen no 
   zero-frequency modes except those associated with rattlers or with 
   uniform translations. 
   This suggests that Point J 
   has no nontrivial zero-frequency modes. 

\begin{figure} 
   \scalebox{0.5}{\includegraphics{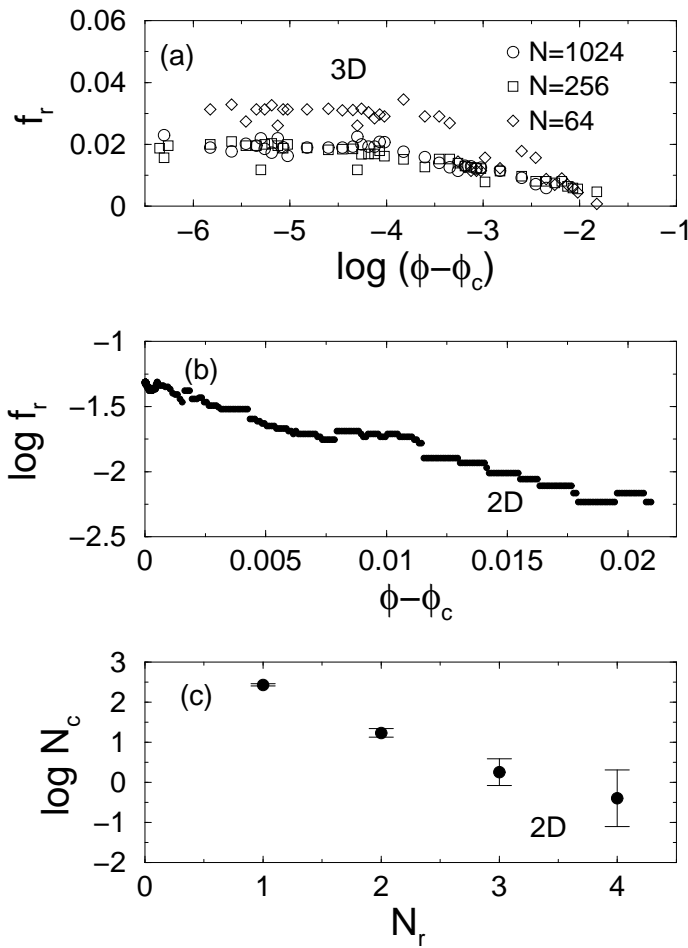}}%
   \caption{(a) Fraction $f_r$ of particles that are rattlers as a
   function of $\phi-\phi_{c}$ for a 3d monodisperse system with harmonic
   repulsions at several system sizes $N$.  (b) $f_r$ versus
   $\phi-\phi_c$ for a $N=1024$ bidisperse system with harmonic
   repulsions in 2d. (c) Number of clusters $N_c$ containing $N_r$ rattlers
   for five $N=10000$ 2d bidisperse systems with harmonic repulsions 
   at $\phi-\phi_c \approx 10^{-2.5}$.}
   \label{rattlers} 
   \end{figure}

 We have studied the fraction $f_r$ of particles that are rattlers as
 a function of $\phi-\phi_{c}$ for both 2d bidisperse and 3d
 monodisperse systems with harmonic interactions.  We show in
 Figs.~\ref{rattlers} (a) and (b) that the fraction of rattlers
 decreases with increasing packing fraction.  We show in
 Fig.~\ref{rattlers}(a) that the fraction of rattlers is independent
 of system size for $N>64$ in 3d.  For the 2d bidisperse
 system, we have also studied the distribution of rattler cluster
 sizes.  We find that most clusters have a single rattler and larger
 clusters are more rare.  This is shown in Fig.~\ref{rattlers}(c).

    \subsection{$g(r)$ diverges at Point J: a vanishing length scale} 
 \label{divergence} 

   A signature of jamming at point J manifests itself in the pair 
   correlation function $g(r)$.  At this point the particles just begin 
   to touch so an important length scale - the distance between nearest 
   neighbor particles - goes to zero.  This vanishing length scale gives 
   rise to a divergence in $g(r)$ in the form of $\delta$-functions at $r 
   = \sigma_{ij}$, the sum of the radii of neighboring particles.  For 
   simplicity, we will focus on monodisperse 
   systems.  Recall from 
   Sec.~\ref{isostatic_section} that at 
   $\phi_{c}^{-}$, there are no contacts whereas at $\phi_{c}^{+}$, $Z$ 
   jumps to the isostatic value $Z_{c}=2d$.  This discontinuity in Z 
   implies that there must be a $\delta$-function in $g(r)$ just at 
   $\phi_{c}$ and that the area underneath this $\delta$-function must be 
   exactly the coordination number at jamming: $Z_{c}=2d$.  This 
   divergence is distinct from the divergence associated with the 
   power-law increase above the first peak in $g(r)$ (where $g(r) \sim 
   (1-r/\sigma)^{-1/2}$ as $r \rightarrow \sigma^{+}$ \cite{silbert}) since 
   that power law is integrable whereas this one has a nonzero area. 

\begin{figure} 
 \scalebox{0.5}{\includegraphics{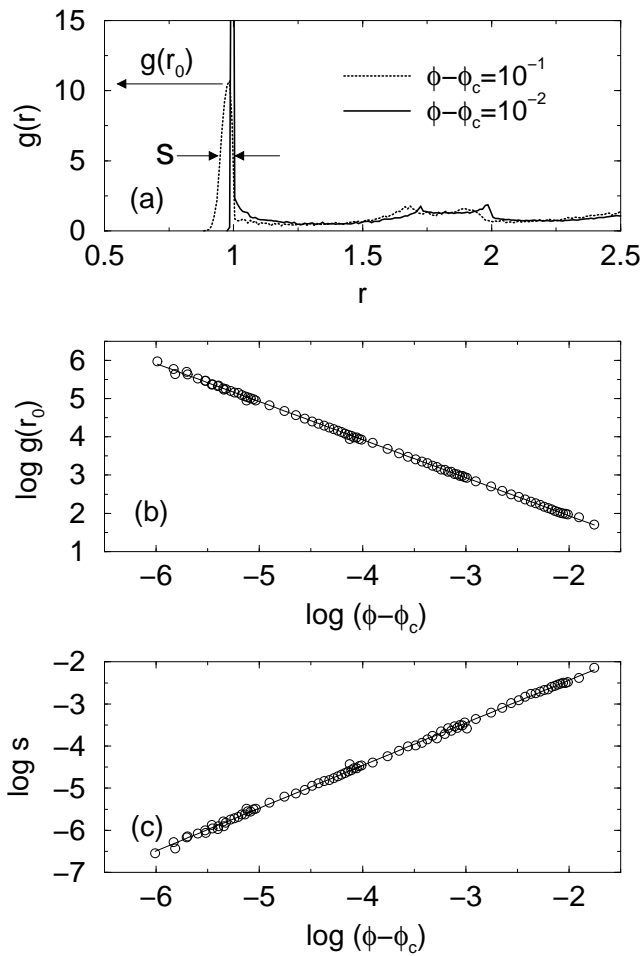}}%
   \caption{(a) The radial distribution function $g(r)$ for a $N=1024$ 
   monodisperse system with harmonic repulsions in  $3d$ at 
   $\phi-\phi_c = 10^{-1}$ and $10^{-2}$.  The height of the first peak 
  $g(r_0)$ and its left-hand-width $s$ are defined. 
 (b) Height of the first peak of $g(r)$ as a function of $\phi-\phi_c$ for 
 the same system as in (a).  The solid line has slope $-1$. 
 (c) Left-hand-width $s$ of the first peak 
   of $g(r)$ as a function of $\phi-\phi_{c}$ for the same system as in (a).
  The solid line has slope $1$.} 
   \label{grphifig} 
   \end{figure}

   Fig.~\ref{grphifig}(a) shows $g(r)$ for a monodisperse, 3-dimensional 
   system at two different values of $\phi-\phi_{c}$.  Note that as $\phi$ 
   approaches $\phi_{c}$ from above, the first peak grows higher and 
   narrower.  We can trace 
   the evolution of the first peak by measuring its height as a function 
   of $\phi-\phi_{c}$ (Fig.~\ref{grphifig}(b)).  We find that the height of 
   the first peak at $r_0$ diverges as a power-law: 
   \begin{equation} 
        g(r_0) = g_{0} (\phi-\phi_{c})^{-\eta} \label{grphi} 
   \end{equation} 
   with $g_0= 0.90 \pm 0.02$ and $\eta = 0.993 \pm 0.002$. 
   Previous hard-sphere 
   simulations\cite{gr_exponent} have measured, with much less 
   precision, the height of the first peak as $\phi_{c}$ is approached 
   from below and found a similar exponent. 

   In Fig.~\ref{grphifig}(c), we plot the left-hand-width at half-height 
   of the first peak of $g(r)$ as a function of $\phi-\phi_{c}$.  This 
   width approaches zero as $\phi \rightarrow \phi_{c}^{+}$ as a 
   power-law: 
   \begin{equation} 
       s = s_0 (\phi-\phi_{c})^{\Delta} \label{grwphi} 
   \end{equation} 
   where $s_0=0.39 \pm 0.04$ and $\Delta=1.01 \pm 0.005$. 

    \subsection{There is an excess low-frequency contribution to the density 
 of vibrational states at Point J} 
 \label{density_of_states_section} 

   The normal modes of vibration provide a complete basis set with which 
   to describe the motions of the particles in a jammed system.  There have 
   been many studies of normal 
   modes in disordered systems \cite{Bell_Dean,rahman,nagel1,
   john,seeley,schober,tanguy,lacks}. 
   In this 
   section we describe the normal mode spectrum as a function of packing 
   fraction above $\phi_c$.  A zero-frequency mode would indicate that 
   some, possibly complicated, set of cooperative displacements of the 
   particles could be made with no cost in energy.  There should always 
   be $d$ such modes corresponding to the simple uniform translation of 
   the system for each of the $d$ dimensions.  Every ``rattler'' 
   particle will likewise contribute $d$ zero-frequency 
   modes.  If a configuration at $\phi = \phi_c$ is isostatic, as we 
   claimed in Section ~\ref{isostatic_section}, then above $\phi_{c}$ the 
   only zero-frequency modes should be the trivial uniform 
   translations of the 
   entire system and of the rattlers.  As we mentioned 
   above we have found no other, nontrivial, zero-frequency modes.  On the 
   other hand, we must expect some change in the nature of the 
   low-frequency modes as the packing fraction for a jammed 
   configuration is lowered toward $\phi_c$.  At that point, some 
   extended mode or modes must approach zero frequency since it is 
   precisely at $\phi_c$ that the system ``falls apart" and becomes 
   unjammed with $dN$ zero frequency modes.  How does the density of 
   states evolve as $\phi - \phi_c$ approaches zero?  In 
   order to compute the normal modes and frequencies $\omega$ of the system, 
   we 
   diagonalize the dynamical matrix of the system \cite{numrec}. 
 The eigenvalues are the squares of the frequencies and the 
 eigenvectors are the polarization vectors of the particles in each 
 mode. 

\begin{figure} 
 \scalebox{0.48}{\includegraphics{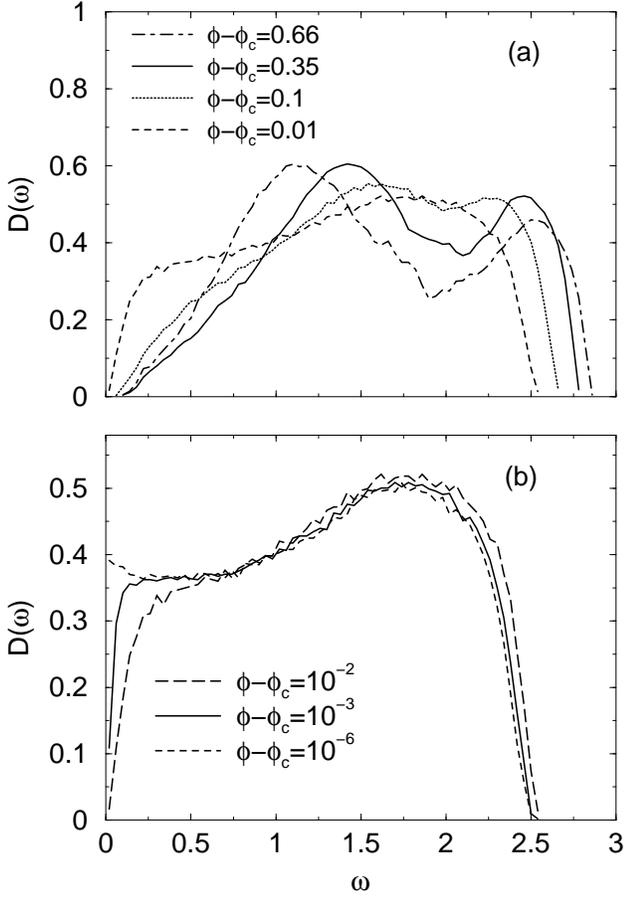}}%
   \caption{Density of states $D(\omega)$ versus $\omega$ for a 3D $N=1024$ 
system with harmonic repulsions at packing
fractions (a) far from $\phi_c$ and (b) close to $\phi_c$.}
   \label{Domega} 
   \end{figure}

 As in a crystal one expects the low-frequency excitations to be the
 long-wavelength sound (longitudinal and transverse) modes.  This
 assumption gives a density of normal mode frequencies, $D(\omega)$,
 proportional to $\omega^{d-1}$.  
 An earlier simulation\cite{nagel1} found
 an increase in the low-frequency density
 of states as the number of nearest neighbors in a glass was reduced.
 As we will show, our present results support this claim.  In the
 previous study \cite{nagel1}, nearest-neighbor bonds were severed
 at random with some probability.  Here, we control the number of
 overlaps by varying the packing fraction, and we can reduce $Z$ all
 the way down to the isostatic value by approaching $\phi_{c}^{+}$.
 In Fig.~\ref{Domega}, we show the density of states obtained for a
 monodisperse harmonic system in 3-dimensions with $N = 1024$
 particles at $T = 0$.  Fig.~\ref{Domega}(a) contains the familiar
 result for compressed systems with packing fractions $\phi$ that are
 far above $\phi_c$.  The largest $\phi-\phi_c$ studied is comparable to 
 typical liquid densities in an equivalent Lennard-Jones
 system\cite{sastry}.  For the very highest packing fractions, we see
 that there is an identifiable region where $D(\omega)$ increases as
 $\omega^2$, as expected.  As the packing fraction is lowered,
 however, we see that the region of $\omega^2$ behavior shrinks,
 reminiscent of the results found in Ref.~\cite{nagel1}.  In
 Fig.~\ref{Domega}(b), we show the behavior of $D(\omega)$ as $\phi$
 approaches $\phi_c$ more closely.  For this 1024-particle system, we
 see no sign of an $\omega^2$ region when $(\phi - \phi_c) \le 0.1$.
 This region has presumably been pushed to low frequencies that are
 inaccessible in a system of this size because the excitations would
 have wavelengths that exceed the linear size of the system.  Even
 though there is no $\omega^2$ behavior at $\phi-\phi_{c}=0.1$,
 Fig.~\ref{Domega}(b) shows that $D(\omega)$ drops as $\omega$ goes to
 zero.  However, as $\phi - \phi_c$ decreases still further, this drop
 in $D(\omega)$ disappears.  By $\phi - \phi_c = 10^{-6}$ there is no
 evidence of it at all and $D(\omega)$ appears to approach a constant
 at zero frequency.  This striking result is unanticipated.  As the
 packing fraction is lowered, the density of states approaches a
 limiting, constant, nonzero value, instead of vanishing as expected
 for long-wavelength sound modes.  Thus, there is a proliferation of
 anomalous low-frequency modes as Point J is approached from above.

   \subsection{Power-law scaling near Point J} 
 \label{power-law_scaling} 

   So far, we have discussed a number of quantities that scale as power 
   laws with $\phi-\phi_{c}$ as the jamming threshold is approached from 
   the high-density side.  Such quantities include the pressure, 
   $p$ 
   (Fig.~\ref{pressure}), the static shear modulus, $G_{\infty}$ 
   (Fig.~\ref{shearmodinfty}) and the coordination number, $Z-Z_{c}$ 
   (Fig.~\ref{Z}).  In 
   addition, we have shown that the width, $w$ (Fig.~\ref{width}), and 
   peak position, $\phi_{0}$ (Fig.~\ref{peak}), of the 
   distribution of jamming thresholds display power-law scaling with 
   system size.  Here we 
   discuss the power-law exponents and their implications. 

 Fig.~\ref{pressure} shows that the pressure 
 vanishes as a power law as $\phi \rightarrow \phi_{c}^{+}$: 
   \begin{equation} 
        p=p_0 (\phi-\phi_{c})^{\psi} \label{ppower} 
   \end{equation} 
 The values for $p_0$ and $\psi$ are listed in 
 Table~\ref{numbers}.  Our results for $\psi$ are consistent with 
 \begin{equation} 
 \psi=\alpha-1, \label{psivalue} 
 \end{equation} 
 independent of polydispersity 
 or dimensionality. 

 The static shear modulus scales as 
   \begin{equation} 
        G_{\infty}=G_{\infty}^{0} (\phi-\phi_{c})^{\gamma} \label{Gpower} 
   \end{equation} 
   where $G_{\infty}^{0}$ and $\gamma$ are listed in Table~\ref{numbers}. 
   The results are consistent with 
   \begin{equation} 
       \gamma=\alpha-3/2, \label{gammavalue} 
   \end{equation} 
   independent of polydispersity or dimensionality. 

 As discussed earlier in Sec.~\ref{isostatic_section}, the coordination 
     number $Z-Z_c$ scales as a power-law with $\phi-\phi_c$ 
     (see Eq.~\ref{Zpower}) with an exponent consistent with 
     \begin{equation} 
     \label{zetavalue} 
         \zeta=1/2, 
     \end{equation} 
     independent of potential, dimensionality, and polydispersity. 
     This result is consistent with 
     earlier estimates from simulations in both $2d$ and $3d$ 
     \cite{ohernJ,durian,makse_packing,silbert}. 

   The height of the first peak of $g(r)$ scales as a power-law: 
   \begin{equation} 
    g(r_{0}) = g_{0} (\phi-\phi_{c})^{-\eta} \label{grhpower} 
   \end{equation} 
   with $\eta = 0.993 \pm 0.002$.  
 This was result obtained for a 3-dimensional monodisperse system with 
 harmonic repulsions.  Similarly, the left-hand width of the first 
 peak of $g(r)$ scales as a power-law: 
   \begin{equation} 
       s = s_0 (\phi-\phi_{c})^{\Delta} \label{grwpower} 
   \end{equation} 
   where $\Delta=1.01 \pm 0.005$.

   Finally, recall the form of the fits to the width and peak position of 
   the jamming threshold distributions, Eqs.~\ref{widthpower} and 
   ~\ref{peakpower}, where the width scales as $w \sim N^{-\Omega}$ and 
   $\phi^{*}-\phi_{0} \sim L^{-1/\nu}$.  Fig.~\ref{width} shows that 
   $\Omega$ appears to be independent of potential, polydispersity and 
   dimensionality.  We find $\Omega=0.55 \pm 0.03$, 
   consistent with 
   \begin{equation} 
       \omega=1/2 \label{Omegavalue}. 
   \end{equation} 
   For the peak position, 
   Fig.~\ref{peakpower} shows that $\nu$ is independent of potential, 
   polydispersity and dimensionality.  We find $\nu= 0.71 \pm 0.08$, 
   consistent with 
   \begin{equation} 
       \nu=2/3. \label{nuvalue} 
 \end{equation} 

   \subsubsection{Interpretation of power-law exponents} 
 \label{interpretation} 

   Some of the exponents for the scalings with $\phi-\phi_{c}$ are 
   straightforward to understand while others are, as yet, without 
   explanation. 

   \subparagraph{Pressure and bulk modulus} 
 \label{pressure_section} 

\begin{figure} 
   \scalebox{0.5}{\includegraphics{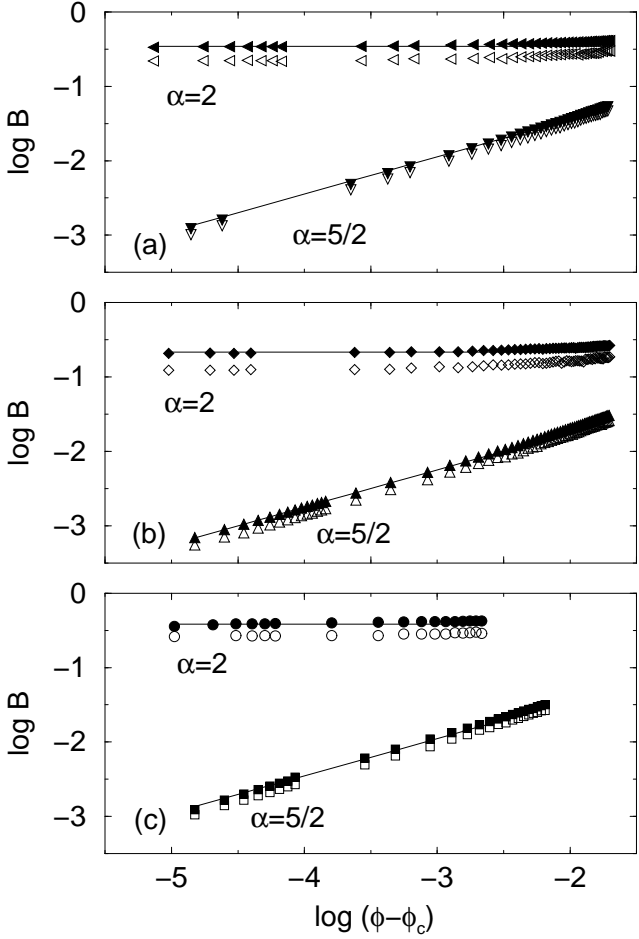}}%
   \caption{Zero-time ($B_{0}$) (closed symbols) and infinite-time 
   ($B_{\infty}$) (open symbols) bulk moduli vs. $\phi-\phi_c$ for 
   (a) 2d bidisperse systems, (b) 3d bidisperse systems, and (c) 3d 
   monodisperse systems with harmonic and Hertzian potentials.  The 
   solid curves have slopes equal to $0$ and $2.0$.  $N=1024$ ($N=512$) 
   particles were used for the $2d$ ($3d$) systems.} 
   \label{bulkmod} 
   \end{figure}

   The exponent for pressure, $\psi \approx \alpha-1$, can be explained 
   if the system responds perfectly affinely to compression. 
   If the deformation 
   is affine, one would expect the exponent for the pressure to be the 
   same as for the force law; this argument yields $\psi=\alpha-1$. 
   Similarly, we would expect the bulk modulus to behave as a power-law: 
   \begin{equation} 
        B \sim (\phi-\phi_{c})^{\beta} \label{Bpower} 
   \end{equation} 
   with $\beta=\alpha-2$ because the bulk modulus is related to the 
   derivative of pressure with respect to packing fraction.  We can 
   check to see if the response of the packing to compression is truly 
   affine by comparing the zero-time bulk modulus, $B_{0}$ to the 
   infinite-time, or static, bulk modulus, $B_{\infty}$.  To obtain 
   $B_{0}$, we apply a compression (or expansion) and measure the change 
   of pressure without allowing any of the particles to relax their 
   positions.  By construction, the compression (expansion) is perfectly 
   affine throughout the sample because we increase (decrease) the radii 
   of all of the particles by the same fixed fraction.  (This is 
   different from how one compresses a sample in a laboratory 
   experiment, where the perturbation is applied at the boundaries of 
   the sample.)  To obtain $B_{\infty}$, on the other hand, we first 
   apply the affine compression (or expansion), then allow the particles 
   to shift their positions by minimizing the energy using the conjugate 
   gradient technique.  If the response to compression is perfectly 
   affine, then the particles will not shift during the conjugate 
   gradient process because the energy is already a minimum.  In that 
   case, we would expect $B_{\infty}=B_{0}$.  The results are shown in 
   Fig.~\ref{bulkmod}.  For all potentials, polydispersities and 
   dimensions studied, we consistently find that $B_{\infty} < B_{0}$, 
   but that they both scale with the same power, consistent with 
   $\beta=\alpha-2$.  These results show that nonaffine deformations due 
   to disorder in the packing do reduce the coefficient of the scaling 
   of the bulk modulus, 
   but do not change the exponent.  It is not obvious why the exponent is 
   unchanged. 

   \subparagraph{Shear modulus} 
 \label{shear_modulus} 

\begin{figure} 
   \scalebox{0.48}{\includegraphics{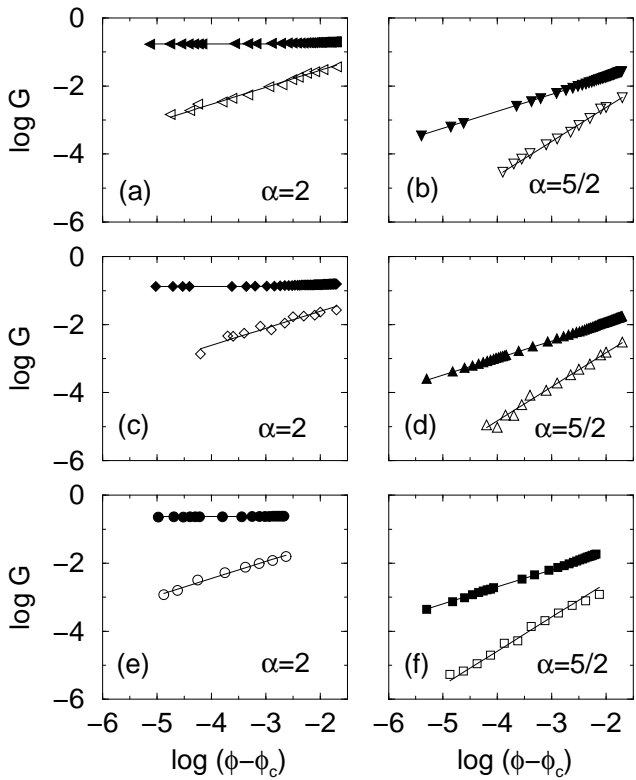}}%
   \caption{Zero-time ($G_{0}$) (filled symbols) and infinite-time 
   ($G_{\infty}$) (open symbols) shear moduli vs. $\phi-\phi_c$ for 
   (a) and (b) 2d bidisperse systems, (c) and (d) 
   3d bidisperse systems, and (e) and (f) 3d monodisperse systems.  Harmonic 
   and Hertzian repulsions are labeled $\alpha=2$ and $\alpha=5/2$, 
   respectively. 
   In (a), (c), and (e) the solid curves have 
   slopes equal to $0$ and $0.5$.  In (b), (d), and (f) the solid curves 
   have slopes equal to $0.5$ and $1.0$.  $N=1024$ ($N=512$) were used 
   for the $2d$ ($3d$) systems.} 
   \label{shearmod0} 
   \end{figure}

   Like the bulk modulus, the shear modulus is also given by two 
   derivatives of the energy.  However, we do not find that the scaling 
   exponent for the static shear modulus, $\gamma$, satisfies 
   $\gamma=\alpha-2$.  Rather, we find $\gamma \approx \alpha-1.5$ (see 
   Eq.~\ref{gammavalue}). 
   To gain insight into this discrepancy, we 
   have examined the zero-time shear modulus, $G_{0}$, as well as the 
   static or infinite-time shear modulus, $G_{\infty}$.  As with the 
   bulk modulus, to measure $G_{0}$ we first apply an affine shear strain and 
   measure the resulting 
   stress without allowing any of the particles to shift their positions. 
   To measure $G_{\infty}$, on the other hand, we apply the conjugate 
   gradient technique once the affine shear is applied and measure the 
   resulting stress {\it after} the energy has been minimized.  Since the 
   shear modulus is the second derivative of the energy, we would expect 
   the exponent for $G_{0}$ to be $\gamma_{0}=\alpha-2$.  This is indeed 
   what we find, as shown in Fig.~\ref{shearmod0}.  The figure shows that 
   $G_{\infty} < G_{0}$, as expected; the 
   system relaxes to a lower value of the shear stress than it has 
   initially.  Although Lacasse, {\it et al.}\cite{mason} have previously 
   pointed out that non-affine deformations can reduce the shear modulus 
   in emulsions, they did not show that the effect of those deformations 
   would be to produce a power-law dependence of the shear modulus upon 
   compression.  Our results show that such power-law scaling exists for 
   the static shear modulus and that the effect of the non-affine 
   deformations is to shift the value of the exponent from 
   $\gamma=\alpha-2$ (appropriate to the $t=0$, affine situation) to 
   $\gamma \approx \alpha-3/2$ (appropriate to the $t=\infty$ case where all 
   relaxation has been allowed to take place).   The 
   effect of non-affine deformation is much more pronounced for the shear 
   modulus than it is for the bulk modulus.  In the latter case, the 
   power-law exponent remained unaffected and only the prefactor was 
   changed.  In the case of the shear modulus, the non-affine deformation 
   changes the scaling exponent as well as the prefactor. 
   As the critical density $\phi_{c}$ is 
   approached from above, the non-affine deformations play a larger and 
   larger role so that $G_{0}/G_{\infty}$ diverges at $\phi_{c}$.

   \subparagraph{Coordination number} 

   Fig.~\ref{Z} shows that the coordination number scales as $Z-Z_{c} 
   \sim (\phi-\phi_{c})^{\zeta}$, where $\zeta$ is independent of 
   potential, polydispersity and dimensionality.  The fact that $\zeta$ 
   is independent of potential is intriguing because it suggests that 
   $\zeta$ depends only on the geometry of the packing.  The fact that 
   $\zeta$ is also independent of dimensionality suggests that there is 
   a property of the packing that is independent of $d$. 

   Recent results\cite{silbert} for the pair correlation function, $g(r)$, 
   of 3-dimensional harmonic packings slightly below the jamming 
   threshold show that $g(r)$ contains a power-law region near 
   $r=\sigma$, where $\sigma$ is the sphere diameter: 
   \begin{equation} 
        g(r) \propto (1-r/\sigma)^{-1/2} \label{grpower} 
   \end{equation} 
   If one assumes an affine deformation upon compression, consistent with 
   the scaling results for pressure and bulk modulus, then one 
   consequence of Eq.~\ref{grpower} is that the 
   coordination number should increase with the power 
   $\zeta=1/2$, as we have observed.  Thus, the scaling in Eq.~\ref{grpower} 
   is consistent with our result $\zeta=1/2$.  The origin of both results, 
   however, is still not understood. 

   \subparagraph{Height and width of first peak of $g(r)$} 

   We find that the height of the first peak of $g(r)$ diverges with an 
   exponent $\eta \approx 1$ (see Eq.~\ref{grhpower}) and that the 
   left-hand-width of the first peak vanishes with an exponent $\Delta 
   \approx 1$ (see Eq.~\ref{grwpower}) as $\phi \rightarrow \phi^+_{c}$. 
   The fact that $\eta \approx \Delta$ is consistent with our 
   expectation that the area of the first peak is roughly $Z_{c}$. 

 \subsubsection{Discussion of finite-size scaling exponents} 
 \label{discussion} 

   We have found that there are very strong system-size effects.  As $N$ 
   diverges, the width of the distribution of jamming thresholds 
   vanishes as $N^{-\Omega}$, leaving a 
   $\delta$-function distribution at Point J.  We find that $\Omega$ is 
   very close to 
   $1/2$ (see Eq.~\ref{Omegavalue}).  It is not obvious that this result 
   can be explained by a simple central limit theorem argument because 
   the packing density is a subtle property of the packing geometry. 
   Independent of the explanation for this exponent, 
   there are still correlations extending across the entire system once 
   it is jammed. 

   The peak 
   position shifts toward the random-close-packing density as $L^{-1/\nu}$. 
   This result suggests that there 
   is a long length scale appearing in the problem near the onset of 
   jamming, which scales as $(\phi-\phi_{c})^{-\nu}$.  Note 
   that our result $\nu=0.71 \pm 0.08$ is a typical value for a 
   correlation length exponent. 

   \subsection{Lack of Self-Averaging at Point J} 
 \label{self-averaging} 

   At Point J, there is no self-averaging in the sense that the average 
   properties of a very large system are not the same as the average 
   over an ensemble of many smaller systems at the same packing 
   fraction.  This property can be understood by considering a system of 
   size $N$ and the behavior as $N$ diverges.  For a finite-sized 
   system, Fig.~\ref{distribution} shows that there is a distribution of jamming 
   thresholds, $\phi_{c}$.  Consider a given packing fraction, $\phi$, 
   that is within this distribution.  Some of the configurations at this 
   $\phi$ will be jammed, and others will be unjammed with $p=0$. 
   For an unjammed configuration, $p=0$ for every subregion of the 
   configuration, as well.  (This is exact even in the infinite 
   system-size limit.)  However, at the same $\phi$ there will exist 
   jammed configurations for which $p>0$.  For those configurations, we 
   have found $p>0$ for almost all subregions.  There are only small 
   clusters of rattlers that have zero local pressures.  The number of 
   such clusters decreases rapidly with the size of the 
   cluster (see Fig.~\ref{rattlers}(c)).   Thus, the value of 
   the pressure averaged over {\it all} configurations cannot be the 
   same as the value of the pressure averaged over an arbitrary given 
   configuration.  As a result, there is no self-averaging.  
   As the system size 
   $N$ increases, the distribution of 
   jamming thresholds narrows.  As a result, the lack of self-averaging 
   will be observed over a smaller region of $\phi$ that eventually 
   narrows to a point (Point J) in the infinite $N$ limit.

   \begin{figure} 
 \scalebox{0.5}{\includegraphics{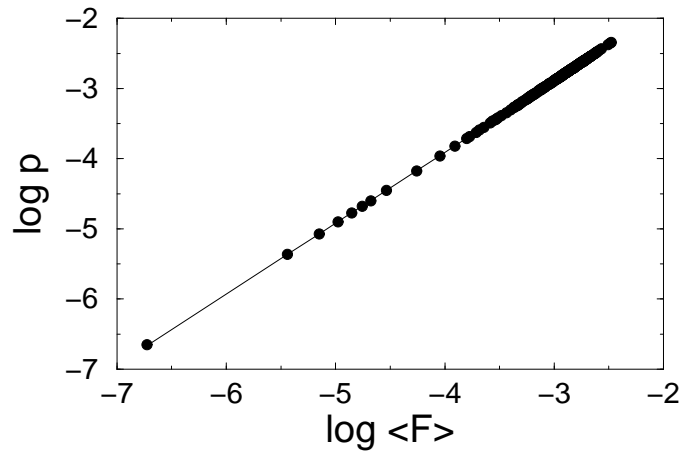}}%
   \caption{Pressure $p$ vs. average interparticle force 
   $\langle F\rangle$ for a $3d$ monodisperse system ($N=512$) with harmonic 
   repulsions. The solid line has slope equal to $1$.} 
   \label{forcepres} 
   \end{figure} 

\begin{figure} 
 \scalebox{0.5}{\includegraphics{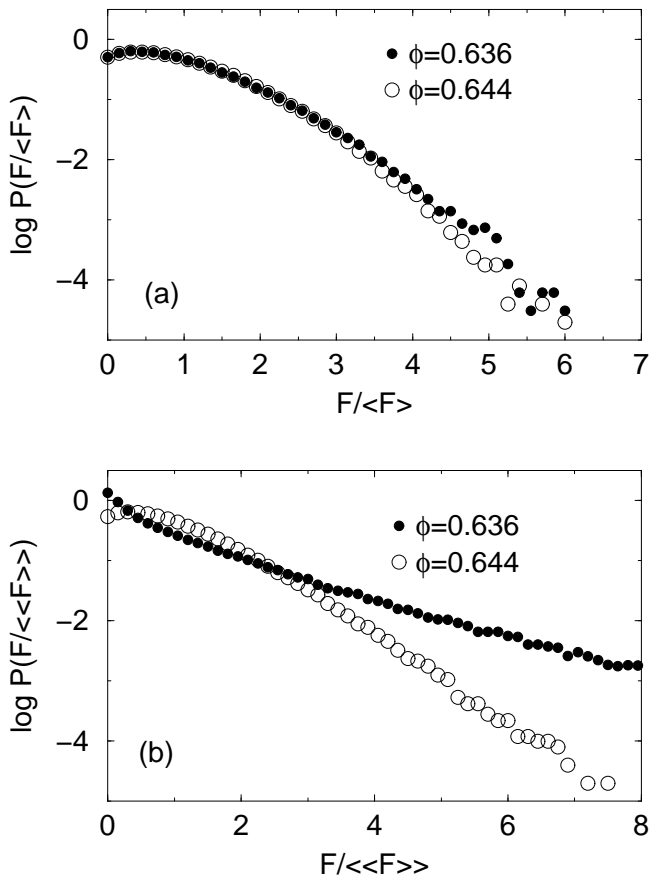}}%
   \caption{Distribution of inter-particle normal forces for a $3d$ 
   monodisperse ($N=1024$) system with harmonic repulsions.  (a) 
   $P(F/\langle F \rangle)$ vs. $F/\langle F \rangle$, and (b) 
   $P(F/\langle\langle F \rangle\rangle)$ vs. $F/\langle\langle F 
 \rangle\rangle$. 
   } 
   \label{forcedist} 
   \end{figure}

   The lack of self-averaging is evident in the distribution of 
   inter-particle normal forces between particles, $P(F)$ \cite{ohernJ}. 
   For a given configuration, the average interparticle force, 
   $\langle F \rangle$ is directly 
   proportional to the pressure of that configuration as shown in 
   Fig.~\ref{forcepres} for a 3d monodisperse system with harmonic 
   repulsions.  Depending on whether one normalizes the forces in a 
   given configuration to $\langle F \rangle$, the average within that 
   configuration, and then averages $P(F/\langle F \rangle)$ over many 
   configurations, or whether one normalizes the forces of all 
   configurations to the same global average force $\langle \langle F 
   \rangle \rangle$, and then calculates $P(F/\langle \langle F \rangle 
   \rangle)$, one will get a different distribution function.  This is 
   shown in Fig.~\ref{forcedist} for a $3d$ monodisperse 
   system with harmonic repulsions.  Note that the 
   difference between $P(F/\langle F \rangle)$ and $P(F/\langle \langle 
   F \rangle \rangle)$ is largest near $\phi=0.636$, which is near the peak, 
   $\phi_{0}$, of the distribution of jamming thresholds for the 
   $3$-dimensional system shown ($N=1024$).  As the packing fraction is 
   increased above $\phi_{0}$, the curves for $P(F/\langle F \rangle)$ 
   and $P(F/\langle \langle F \rangle \rangle)$ look more and more 
   similar.  This is consistent with the argument above, that the lack 
   of self-averaging is most pronounced near the peak of the 
   distribution of jamming thresholds.  A simple argument for the shape 
   of the tail of $P(F/\langle \langle F \rangle \rangle)$ was given 
   earlier \cite{ohernJ}.

   \subsection{Critical behavior near Point J} 
 \label{critical_behavior} 

   In many ways, point J resembles a critical point.  We have shown in 
   Figs.~\ref{pressure}, ~\ref{shearmodinfty}, ~\ref{Z}, ~\ref{grphifig},
   ~\ref{bulkmod}, and ~\ref{shearmod0} that there is 
   power-law scaling near Point J of 
   quantities such as the pressure, shear modulus, bulk modulus, 
   coordination number, and the height and width of the first peak of 
   the pair correlation function\cite{ohernJ}.  We have also shown 
   in Figs.~\ref{width} and \ref{peak} that there is finite-size scaling 
   since the width and peak position of the distributions of jamming 
   thresholds scale with the size of the system.  This is reminiscent 
   of behavior near an ordinary critical point.  Finally, we 
   demonstrated in Fig.~\ref{forcedist} that properties such as the force 
   distribution do not self-average near Point J.  As the system size 
   increases, the packing fraction must be tuned closer and closer to 
   the peak of the distribution of jamming thresholds in order to see 
   the breakdown of self-averaging.  This is also what one expects near 
   an ordinary critical point, where the temperature must be tuned closer and 
   closer to the critical point as 
   the system size increases in order for the correlation length to 
   exceed the system size. 

\begin{figure} 
 \scalebox{1.0}{\includegraphics{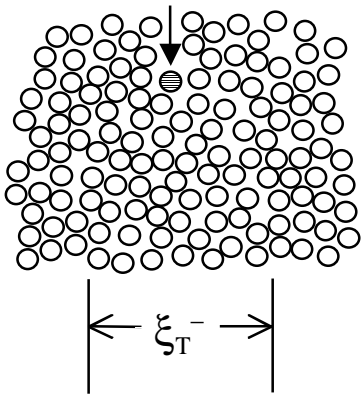}}%
   \caption{Sketch of the transverse length scale.} 
   \label{transverse} 
   \end{figure}

   The lack of self-averaging near Point J and the power-law scaling of 
   the width and peak position of the jamming threshold distribution 
   with system size all suggest that there is a correlation length that 
   diverges at Point J.  What might this length scale be?  We speculate 
   that there is a transverse length scale that does diverge as point J 
   is approached from below.  If the system 
   is held at a packing fraction slightly below the critical value, 
   the system is unjammed and the particles can all move and rearrange. 
   However, the number of particles that must move in order to allow a 
   rearrangement will depend on how close one is to the transition.  Thus, 
   in an infinite system, if one applies a fixed, infinitesimal velocity 
   to a single particle we would expect the particle to disturb the 
   surrounding particles as it moves.  This disturbance will extend to a 
   distance $\xi^{-}_{T}$, the transverse length scale, in a direction 
   perpendicular to the applied force.  We expect that $\xi^{-}_{T}$ will 
   diverge as one gets close to the transition because as the density 
   approaches the close packing value, more and more particles must rearrange 
   to allow for the single particle motion in the longitudinal direction. 
   The idea behind this transverse length scale is shown in 
   Fig.~\ref{transverse}.  Similar ideas are currently being explored 
   experimentally in granular systems with friction \cite{clement} and 
   in colloidal systems~\cite{weeks,reichardt}. 

   One might estimate the transverse length scale by computing how many 
   particles must move laterally in order to insert an extra particle. 
   This is the parking lot model \cite{parking}.  According to this 
   argument, the transverse length scale should diverge as $\xi^{-}_{T} 
   \sim (\phi_{c}-\phi)^{-1/(d-1)}$, for $\phi<\phi_{c}$, where $d$ is 
   the dimensionality.  We note, however, that this result does not 
   agree with the correlation length exponent that 
   we obtained from the finite-size scaling analysis (see 
   Eq.~\ref{nuvalue}), which appears to be independent of dimensionality.

   Although Point J resembles a critical point, it has properties unlike 
   any other critical point ever studied.  The exponents appearing in 
   the scaling relations are independent of dimension but do depend on 
   the potential.  The former observation could be reconciled with a 
   normal critical point if the upper critical dimension for jamming 
   were less than 2, but then we would not expect different potentials 
   to yield different exponents.  Likewise, if each different potential 
   were in a different universality class and yielded different 
   exponents, then the upper critical dimension should be above 3. 
   There are other properties of Point J that are unusual (although not 
   unheard of) for a critical point.  At packing fractions below Point 
   J, the pressure, shear modulus and contact number are all zero and the 
   energy is zero everywhere.  There are 
   no fluctuations in these quantities, even infinitesimally close to Point J as 
   $\phi 
   \rightarrow \phi^{*}$ from below.  In addition, there is a 
   discontinuous jump in the value of the coordination number $Z$ from 
   zero to $Z_{c}$ at $\phi^{*}$.  We also note that we have identified 
   a length scale that goes to {\it zero} at this point:  the spacing 
   between particles that form the connected network in the jammed 
   state.  This is seen in the divergence of the first peak of $g(r)$ 
   (Fig.~\ref{grphifig}).  At a critical point, one expects a single 
   divergent lengthscale and not a lengthscale going to zero.

   Perhaps the most disturbing feature of Point J, from the point of view 
   of ordinary critical phenomena, is the difference in the behavior at 
   fixed pressure and fixed volume.  At fixed volume, we observe 
   finite-size rounding of power-law scaling and finite-size effects 
   such as the lack of self-averaging.  This is because different 
   states have different jamming thresholds, $\phi_{c}$.  At a fixed 
   $\phi$, different states are averaged together and the clean 
   power-law behavior we observe as a function of $\phi-\phi_{c}$ will 
   be rounded.  However, Fig.~\ref{pressure} shows that a fixed pressure 
   corresponds to a fixed value of $\phi-\phi_{c}$.  When we plot 
   quantities as a function of $\phi-\phi_{c}$, we do not see finite-size 
   rounding of power-law behavior.  Even for a finite-sized system, the 
   behavior of the shear modulus, etc. appears to be a clean power-law 
   down to the smallest measurable values of $\phi-\phi_{c}$.  Thus, we 
   do not see measurable finite-size effects at fixed pressure.  The divergence 
   of $g(r)$ also occurs even for a finite-sized system.  These results 
   are very different from what one would expect for an ordinary 
   critical point.

   \section{Implications of Point J for the glass transition} 
 \label{implications} 

   We have shown that Point J marks a well-defined transition from the 
   unjammed to the 
   jammed state.  Because the conjugate gradient method allows us to 
   probe the infinite-time behavior of the system, we have been able to 
   show that the system develops a truly static shear modulus at Point 
   J.  Where Point J lies with respect to the jamming surface depicted 
   in Fig.~\ref{jpd} depends on one's definition.  Since the glass 
   transition line is usually defined as the temperature where the 
   relaxation reaches some large but finite threshold value, Point J in 
   this definition 
   strictly lies within the jammed phase since the relaxation time there 
   is infinite.  Since Point J lies just below the jamming surface of 
   the phase diagram, one might expect it to control behavior in its 
   vicinity if it is indeed a critical point.  If so, it may be the 
   long sought-after phase transition underlying the glass transition. 
   In this section, we discuss why we 
   suspect that the physics of point J may hold clues for understanding 
   the entire jamming surface of Fig.~\ref{jpd}, including the glass 
   transition itself. 

\begin{figure} 
 \scalebox{0.9}{\includegraphics{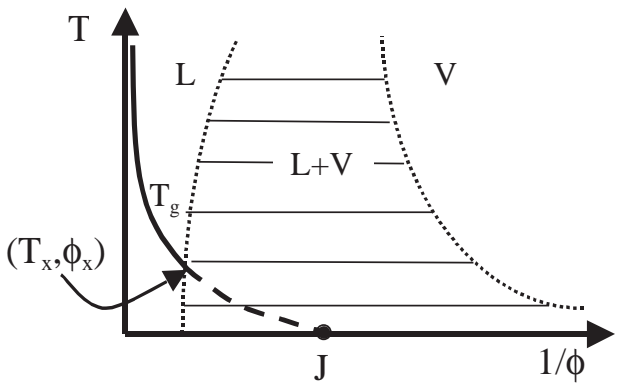}}%
   \caption{A sketch of the jamming phase diagram in the $T-1/\phi$ plane 
   for a system with 
   short-ranged repulsion and longer-ranged attraction.  For 
   simplicity, we have 
   assumed that crystallization does not occur.  The jammed 
   region lies underneath the curve marked $T_{g}$.  Beyond 
   $(T_{x},1/\phi_{x})$, where the glass transition curve crosses the 
   liquid density at coexistence, the available states depend on quench 
   history.} 
   \label{jpd2} 
   \end{figure}

   One might 
   wonder why Point J is important to real glass-forming liquids, 
   where there are 
   not only finite-ranged repulsive 
   interactions such as those we have included in our calculations, but 
   also longer-ranged attractions. 
   The jamming phase diagram for a 
   real liquid would look quite different from the one depicted in 
   Fig.~\ref{jpd}.  In addition to the jamming surface, one has to 
   consider the vapor-liquid phase coexistence curve once particles can 
   attract one another.  In 
   Fig.~\ref{jpd2}, we have sketched the jamming phase diagram in the $T-1/\phi$ 
   plane when attractions are present.  For simplicity, we have 
   explicitly assumed that there is no possibility of 
   crystallization.  (If crystallization were taken into account, then 
   the liquid that coexists with vapor could be metastable to the crystal.) 
   In Fig.~\ref{jpd2}, the glass transition temperature 
   decreases with increasing $1/\phi$ and eventually crosses the 
   liquid-vapor coexistence region at $(T_{x},1/\phi_{x})$, as shown.  Once the 
   glass transition 
   curve crosses the left-hand side of the coexistence curve, which 
   represents the lowest accessible liquid density, a variety of states 
   can be obtained 
   depending on the quench history.  The dashed 
   part of the glass transition curve, which ends at Point J, is not 
   necessarily accessible to systems with liquid-vapor phase transitions.

   Even though Point J does not necessarily exist for real liquids, 
   it can still influence the glass transition.  In systems 
   with short-ranged repulsions and longer-ranged attractions, there is 
   still a well-defined distance at which the repulsion vanishes; this is 
   the position of the minimum in the pair potential.  As with 
   the theory of liquids, attractions are a small perturbation to the 
   strong repulsive core; they merely hold the system at a sufficiently 
   high density that the repulsions can come into play\cite{wca}.  We 
   therefore expect the behavior we find near Point J to be a good 
   approximation to the behavior of liquids down to the density at which 
   the glass transition line crosses the liquid-vapor coexistence curve. 

   \subsection{Significance of divergence in pair correlation function} 
 \label{pf} 

 We noted above that the first peak of the pair correlation function 
 $g(r)$ diverges at Point J.  This has two consequences that have been 
 observed in studies of the glass transition.  The first has to do with 
 the static structure factor, $S(k)$, measured from scattering 
 experiments, and the second has to do with the emergence of a peak in 
 the distribution of normal forces, $P(F)$, as measured experimentally 
 in granular\cite{3dexppf} and colloidal\cite{brujic} systems, and 
 numerically in previous work on models of 
 glass-forming liquids\cite{ohernpf}. 

\begin{figure} 
 \scalebox{0.5}{\includegraphics{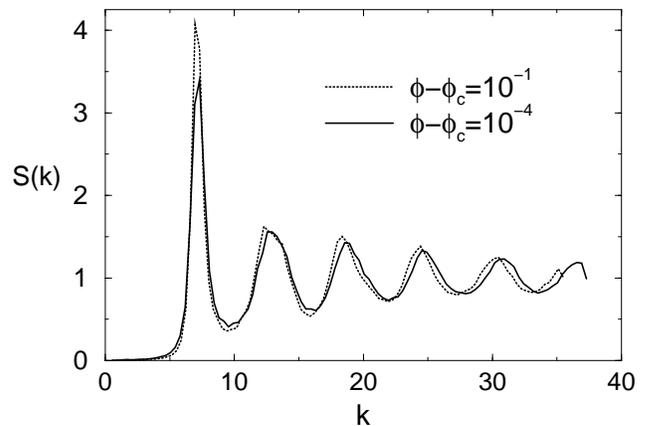}}%
   \caption{Static structure factor $S(k)$ at $\phi-\phi_c=10^{-1}$ and 
   $10^{-4}$ for a 3d monodisperse system with harmonic repulsions.} 
   \label{Sk} 
   \end{figure}

 At Point J, the first peak 
 of $g(r)$ is infinitely high and narrow.  This property 
 elucidates one heretofore puzzling aspect 
   of studies of supercooled liquids.  The static structure factor, 
   $S(k)$, is related by a Fourier transform to $g(r)$ so the 
   $\delta$-function peak in $g(r)$ produces oscillations in $S(k)$. There 
   will not be a divergence in $S(k)$ at any wavevector $k$.  This is 
   different from what one finds at a critical point where there is a 
   diverging susceptibility at some value of $k$.  (In the case of a 
   ferromagnetic transition, this would be the magnetic susceptibility 
   at $k = 0$.)  Thus, the signature of the transition at point J is 
   different from that observed in ordinary second-order phase 
   transitions.  As one moves away from Point J into the jammed region, 
   the $\delta$-function in $g(r)$ broadens and decreases in height, but the 
   oscillations in $S(k)$ persist.
   Representative plots of $S(k)$ at two 
   different values of $\phi-\phi_{c}$ are shown in Fig.~\ref{Sk}. 
   These are qualitatively similar to experimental results for $S(k)$ at 
   high $k$, which also show oscillations\cite{exptSk}.  This 
   clarifies why searches for structural signatures of 
   the glass transition that have examined the shape of $S(k)$, either 
   at small $k$ 
   or in the vicinity of the first peak, 
   have not found divergent behavior. 

     It has long been recognized that the first peak of $g(r)$ rises and 
    sharpens as the temperature is lowered toward the glass transition. 
    However, the change of behavior as one crosses the glass transition 
    is only quantitative.  A criterion suggested many years 
    ago\cite{gr_peak}, that the glass transition occurs when the first 
    peak reaches a threshold height, seems rather arbitrary.  In a 
    previous study \cite{ohernpf} we showed that there is a {\it 
    qualitative} change in a quantity closely-related to $g(r)$.  This 
    is the distribution of normal forces, $P(F)$: 
     \begin{equation} 
        P(F) dF \propto r^{d-1} g(r) dr \label{pfgr} 
   \end{equation} 
   where $d$ is the dimensionality of the system.   This quantity has 
   been measured experimentally at the boundaries of static granular 
   packings \cite{3dexppf} and in the interior of colloidal glasses 
   \cite{brujic}.  In all these studies of jammed systems, 
   $P(F)$ was found to contain a peak.  Our previous studies show that a 
   peak develops in $P(F)$ as the jamming surface is approached by 
   lowering $T$, 
    increasing $\phi$, or decreasing $\Sigma$ \cite{ohernpf,silbert2}.  
    This signature 
    was observed for 
    all the potentials we have studied, including the full Lennard-Jones 
    interaction, the Weeks-Chandler-Andersen (WCA) 
    interaction\cite{wca}, harmonic repulsions, and Hertzian repulsions. 
    Thus, the development of a peak in $P(F)$ provides a signature of the 
    onset of jamming from purely 
    structural data. 
    From Eq.~\ref{pfgr}, one can show that $P(F)$ develops a peak only if the 
   first peak of $g(r)$ is sufficiently high and narrow.  The criterion for 
   a peak in $P(F)$ is 
    \begin{equation} 
   \frac{ d\ln g}{dr} = \frac{1-d}{r} + \frac{d^2 F/dr^2}{dF/dr}
   \label{pfgrpk} 
   \end{equation} 
   The fact that the onset of jamming is correlated with the first peak 
   of $g(r)$ becoming high and narrow enough suggests that the entire 
   jamming surface may be controlled by 
   Point J. 

\begin{figure} 
 \scalebox{0.48}{\includegraphics{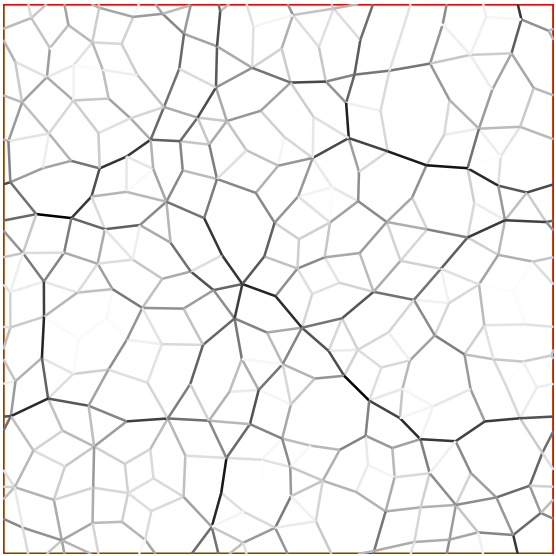}}%
   \caption{Network of interparticle forces for a $2d$ bidisperse 
   system with harmonic repulsions at $\phi-\phi_c=10^{-4.5}$ and $N=256$. 
   The intensity of the line shading is proportional to the magnitude of the 
   interparticle force.} 
   \label{network} 
   \end{figure}

   In order for a 
   system to jam, it must be able to support shear stress for a very long 
   time.  The stress is supported through a network of inter-particle forces, 
   suggesting that an order parameter 
   for jamming may be found in the nature of such a network. 
   Forces on a particle must either be balanced by other forces or give 
   rise to accelerations.   At high temperatures, there is a lot of 
   kinetic energy and particles are constantly accelerated by unbalanced forces. 
   At lower temperatures, however, 
   the forces on particles tend to balance more because accelerations 
   are smaller, and at zero 
   temperature, forces on particles 
   balance perfectly so that the system is mechanically stable at packing 
   fractions above Point J.  The 
   resulting network of forces at $T=0$ is shown just above the onset of 
   jamming in Fig.~\ref{network}.  The order parameter for the glass 
   transition presumably depends on 
   at least a three-particle quantity in order to characterize the force 
   network.   However, $P(F)$, 
   which is only a two-particle quantity, clearly couples to the force 
   network.  A peak in $P(F)$ reflects 
   the existence of the network because the forces on all particles can only 
   balance if they are of roughly the same magnitude.  This intuition highlights 
   the importance of Point J to the glass transition. 
   At $T=0$, as the packing fraction is 
   increased through Point J, the number of overlaps jumps from $Z=0$ (no 
   force network) to 
   $Z=2d$ (a dense force network as shown in Fig.~\ref{network}).  Thus 
   Point J marks the development of a force network that supports shear 
   stress. 

    \subsection{Significance of anomalous low-frequency modes in density 
    of states} 
 \label{dos} 

 Perhaps the most striking evidence that the physics at 
 Point J may be related to the nature of glasses and the glass transition 
 is to be found in the behavior of the density 
 of vibrational states at low frequencies.  In contrast to our 
 expectation that the density of states should vary as 
 $D(\omega) \propto \omega^2$ at low frequencies (in 3-dimensions), we 
 find that at Point 
 J, the density of states approaches a non-zero constant value as 
 $\omega \rightarrow 0$ (see Sec.~\ref{density_of_states_section}). 

 We suspect that these extra low-frequency modes are primarily 
 transverse in nature. 
 It is clear from the behavior of the zero- and infinite-time shear moduli, 
 $G_{0}$ and $G_{\infty}$, that 
 the transverse modes must become increasingly soft due to the 
 relaxation allowed by non-affine deformations as $\phi$ 
 approaches $\phi_c$.  The ratio, $G_{0}/G_{\infty}$, diverges 
 at this point (see Sec.~\ref{shear_modulus}).  The bulk 
 modulus, in contrast, does not show any particular softening 
 due to non-affine relaxations and $B_{0}/B_{\infty}$, is a 
 constant as $\phi$ approaches $\phi_c$ (see Sec.~\ref{pressure_section}). 
 This suggests that the anomalous 
 low-frequency modes are more transverse than longitudinal in character. 
 Moreover, since $G_{0}/G_{\infty}$ diverges, and the difference 
 between $G_{0}$ and $G_{\infty}$ arises from spatially inhomogeneous 
 non-affine relaxations, we expect that 
 there must be significant high-wavevector contributions mixed 
 into the anomalous modes \cite{nagel2}. 

 As $\phi \rightarrow \phi_{c}^{+}$, we 
 also know that the normal modes are becoming more anharmonic. 
 This was shown in Fig.~\ref{shearmod} where it is clear that the linear 
 region of the stress versus strain curves becomes smaller as 
 $\phi_c$ is approached.  The effect of this anharmonicity 
 still needs to be determined. 

 Our results, that anomalous low-frequency 
 vibrational modes proliferate and herald unjamming as $\phi$ 
 approaches Point J, are of clear relevance to a large body of 
 experimental data on excess vibrational modes in glasses.  Two 
 results reflect these excess states rather directly. 
 The first is the boson 
 peak, measured by light and x-ray scattering\cite{bosonexp} and 
 in simulations\cite{bosontheory}, which 
 indicates an 
 excess of vibrational states at low 
 frequencies, above those predicted by Debye behavior ($D(\omega) 
 \propto \omega^{2}$ in three dimensions).   The second is the 
 low-temperature specific heat of glasses: 
 \begin{equation} 
      c_v = A_{\rm Debye} T^3 + B T + C_{\rm excess}T^3 
 \end{equation} 
 In addition to the Debye term from long-wavelength sound modes, 
 there is a linear term in the specific heat and an excess 
 $T^3$ term above that predicted by the velocities of sound. 
 The linear term has been ascribed to the existence of a new type of 
 mode: two-level tunneling systems \cite{anderson,phillips}. 
 We note that a constant density of states as we have found at 
 Point J, would, by itself, produce a linear term without the 
 necessity of assuming a new set of tunneling excitations. 
 However, since glasses exist well above Point J, we would not expect 
 such a linear term to persist all the way down to zero temperature. 
 Nevertheless, there is still 
 a remarkable excess density of states even far away from Point J 
 which would contribute to both the excess $T^3$ term and to 
 the boson peak--both strong signatures of glassy behavior. 

 \section{Conclusions and Conjectures} 

 We have 
 demonstrated a number of remarkable properties of one special point on 
 the jamming phase diagram that occurs at zero temperature and zero applied 
 shear stress.  We have shown that this Point J is the onset of jamming 
 in both the bulk and shear modulus, it is well defined in the $N 
 \rightarrow \infty$ limit, and provides a clean definition of ``random 
 close-packing.'' In many ways it behaves as a 
 critical point while in others it has properties not normally 
 associated with a second-order phase transition.  For example, many 
 quantities, such as pressure and shear modulus, scale as power laws 
 with 
 $\phi - \phi_c$ but the scaling exponents depend on the inter-particle 
 potential and not on dimension. 

 From the finite-size scaling results 
 that we have reported, one might also conjecture how quenched disorder imposed 
 externally (such as from pinning sites in a flux lattice or from 
 optical traps in a colloidal suspension) would affect the nature of 
 the jamming phase diagram.  If we assume that the spacing between 
 defects limits the correlation length in the system, instead 
 of the finite size of the box that we employed in these studies, then 
 we would expect that the jamming threshold at Point J would be smeared out 
 in much the same way as we find in finite-sized systems. 
 Thus, if we were to add a ``quenched disorder" axis on the jamming 
 phase diagram, one of the implications of our work would be that as 
 more quenched disorder is added, the 
 distribution in jamming thresholds will broaden. 

 Our studies here have been confined to purely frictionless 
 particles.  We suspect that for systems with frictional interactions, 
 the distribution of jamming thresholds should broaden as well.  This 
 would be in accord with experimental observations that static 
 frictional packings can exist over a wide range of densities. 

 Perhaps most significant is that at Point J many of the properties of
 disordered glassy systems have their most pronounced expression.
 Just as a crystal is the most ordered of states, Point J may be
 considered to be the most disordered of states.  As at a critical
 point, where the correlations across the entire system are most
 easily observable, at Point J the nature of the disordered phase is
 most plainly seen.  The constant density of low-frequency normal
 modes and the divergence in the first peak in $g(r)$ are two
 extraordinary examples.  Both are sensitive to global properties of
 the system; $D(\omega)$ because it deals with the longest wavelength
 modes in the system, and $g(r)$ because the overlap between all
 particles simultaneously goes to zero.  In addition, both of these
 observations have implications for how real glassy systems behave.
 It is tempting to think that Point J may provide a key to
 understanding the nature of the entire surface in the jamming phase
 diagram and to argue that the properties of other glassy states
 should be understood as a perturbation around this ``most disordered"
 of states.  Thus, one might say that Point J represents the epitome
 of disorder and the essence of glassiness.

 \acknowledgments We thank Bulbul Chakraborty, Sue Coppersmith,
 Morrell Cohen, Doug Durian, Gary Grest, Leo Kadanoff, Steve Langer,
 Dov Levine, Gilles Tarjus, Denis Weaire and Tom Witten for helpful
 discussions.  We also gratefully acknowledge the support of 
 NSF-DMR-0087349 (AJL) and NSF-DMR-0089081 (SRN).

 \begin{table*} 
 \caption{\label{numbers}Coefficients and exponents for the power-law 
 scaling of pressure $p$, shear modulus $G_{\infty}$, and 
 coordination number $Z-Z_c$ for all systems studied.} 
   \begin{tabular} {|c|c|c|c|c|c|c|c|c|c|} \hline 
 \multicolumn{10}{|c|}{Power-law scaling} \\ \hline 
 \multicolumn{3}{|c|}{System} & \multicolumn{7}{|c|}{Quantity} \\ \cline{4-10} 
   \multicolumn{3}{|c|}{} &  \multicolumn{2}{|c|}{$p$} 
   & \multicolumn{2}{|c|}{$G_{\infty}$} & 
 \multicolumn{3}{|c|}{$Z-Z_c$} \\ \hline 
   d & polydispersity & $\alpha$ & $p_0$ ($\pm 0.05$) & 
 $\psi$ ($\pm 0.03$) 
 & $G^0_{\infty}$ ($\pm 0.05$) & $\gamma$ ($\pm 0.05$) & $Z_0$ ($\pm 0.5$) 
 & $\zeta$ ($\pm 0.04$) & $Z_c$ ($\pm 0.02$)\\ \hline 
   2 & Bi & 2 & 0.34 & $1.01$ & 0.24 & $0.47$ & 3.6 
   & $0.49$ & $3.98$ \\ \hline 
   2 & Bi & 5/2 & 0.27 & $1.50$ & 0.21 & $0.99$ & 3.3 
   & $0.48$ & $3.98$ \\ \hline 
   3 & Bi & 2 & 0.28 & $1.03$ & 0.21 & $0.48$ & 8.4 
   & $0.47$ & $5.98$ \\ \hline 
   3 & Bi & 5/2 & 0.18 & $1.51$ & 0.17 & $1.02$ & 7.4 
   & $0.49$ & $5.98$ \\ \hline 
   3 & Mono & 2 & 0.48 & $1.01$ & 0.34 & $0.49$ & 7.7 
   & $0.51$ & $5.98$ \\ \hline 
   3 & Mono & 5/2 & 0.35 & $1.50$ & 0.14 & $0.95$ & 7.7 
   & $0.47$ & $5.98$ \\ \hline 
   \end{tabular} 
\end{table*}

 \end{document}